\documentclass[final,5p,times,twocolumn]{elsarticle}
\usepackage{amsmath,amssymb,amsfonts,scalerel}
\usepackage{blindtext}
\usepackage{graphicx}
\usepackage{adjustbox}
\usepackage{epsfig}
\usepackage{float}
\usepackage{subfig}
\usepackage{xcolor}
\usepackage[colorlinks=true,citecolor=blue,linkcolor=blue,urlcolor=blue]{hyperref}
\usepackage{setspace}
\usepackage{multirow}
\usepackage{multicol}
\usepackage{url}
\usepackage{siunitx}
\usepackage{hyperref}
\usepackage{rotating}
\usepackage{geometry}
\biboptions{sort,compress}
\journal{Journal of Environmental Sciences}

\begin{document}

\begin{frontmatter}

\title{Temperature dependence study of water dynamics in Fluorohectorite 
       clays using Molecular dynamics simulations}
\author{H.O.~Mohammed}
\ead{hussen.oumer@aau.edu.et}
\author[]{K.N.~Nigussa\corref{cor1}}
\cortext[cor1]{Corresponding author:\ kenate.nemera@aau.edu.et\ 
               (K.N.~Nigussa)}
               
\address{Department of Physics,\ Addis Ababa University,\ 
         P.O. Box 1176,\ Addis Ababa,\ Ethiopia}
                  
\begin{abstract}
In this work, we have carried out molecular\ 
dynamics~(MD) simulation\ techniques to study the\ 
diffusion coefficient\ of interlayer molecules\
at different temperature.\ Within the wider\ 
context of water dynamics\ in soils,\ and with\ 
a particular emphasis\ on clays,\ we present\ 
here the\ translational dynamics\ of water in\ 
clays,\ in a bi-hydrated states.\ We focus\
on temperatures\ between 293~K and 350~K, i.e.,\ 
the range relevant to the\ environmental waste\ 
packages.\ A natural hectorite\ clay of interest\ 
is\ modified as a\ synthetic clay,\ which allows\ 
us to understand\ the determinantal\ parameters\ 
from MD simulations\ through a comparison with\ 
the\ experimental\ values.\
The activation energy\ $E_{a}$ determined by\ 
our simulation is $[8.50-16.62] \frac{kJ}{mol}$.\
The calculated diffusive\ constants are in the\ 
order of 10$\rm^{-5}~cm^{2}s^{-1}$.\ The simulation\ 
results are in good agreement\ with experiments\ 
for the relevant set of\ conditions,\ and they\ 
give more insight\ into the origin of\ the\ 
observed dynamics.\   
\end{abstract}


\end{frontmatter}

\section{Introduction\label{sec:intro}}
Clay~minerals\ are abundant,\ non-toxic,\ cheap\ 
and reusable~\cite{michels2015intercalation}.\ 
They have\ large surface areas,\ making them ideal\ 
for applications\ in which we want\ to adsorb other\ 
molecules on\ the surface.\ Because of\ the layered\ 
nature\ of\ the\ structure of\ these materials,\ 
inserting different\ molecules between\ layers can\ 
result in adsorption\ with high binding\ energies.\ 
Today,\ clay, especially smectite clay,\ are studied\ 
for different applications\ like the possibilities\ 
for a carbon storage~\cite{giesting2012interaction,
michels2015intercalation,galan2014experimental,azzouz2010carbon},\ 
water retention\ and adsorpition~\cite{michels2020impact,
larsen2020physicochemical,altoe2016continuous},\ 
and drug\ delivery mechanism~\cite{dos2017ciprofloxacin,
hernandez2018synthetic,dos2018ph}.\ 
While clay\ mineral deposits\ can be found\ all over the\ 
world,\ their composition\ and impurities vary.\ 
Although\ kaolins and smectites\ are widely available\ 
in the commercial market,\ several useful\ clay minerals\ 
are not\ abundant in nature.\ Furthermore,\ two major\ 
issues arise\ in the use\ of clay minerals.\
These\ are the depletion\ of the natural deposits,\ 
especially those\ with easy access\ for mining,\ 
and the occurrence\ of this clay\ as a mixture of\ 
several phases\ instead of a pure\ single phase.\ 
To overcome\ these problems,\ scientists\ 
in the fields\ of geology,\ material science,\ and\ 
geochemistry\ have been\ taking particular\ interest\ 
in the\ laboratory synthesis\ of clay minerals over\ 
recent decades~\cite{zhou2010synthetic,zhang2010synthesis,
zhou2005structure}.\
In this work,\ we will consider\ a specific type\ 
of clay material\ in the\ Synthetic smectite group,\
called Fluorohectorite.\ It is an\ important material\ 
in clay science\ that definitely\ underlines the\ 
statement that\ "Clays may be\ considered as the\ 
material of the 21$\rm^{st}$ century"~\cite{bergaya2013handbook}.\
Analyzing\ water dynamics\ in anisotropic\ and\ 
confined media\ such as\ clay minerals\ is vital\ 
for comprehending\ transport features\ in such\ 
materials\ as\ split solids,\ porous media,\ soil\ 
models, in environmental sciences.\ Water plays\ 
a crucial role for\ clay materials\ when it comes\ 
to the storage\ of high radioactive\ wastes,\ since\ 
the waste is\ surrounded by water\ and ions that\ 
can disperse\ inside the clay,\ in what is called\ 
a phenomenon\ that requires\ long-term forecasting.\ 
This necessitates\ the\ understanding of\ the\ 
behaviour\ of water on\ several length-scales,\ 
the smallest\ of which\ is the nanoscale,\ and\ 
also for a range\ of temperature\ between\ 293~K\ 
and 350 K.\
\section{Computational Methods}\label{sec:comp}
\subsection{System setup}
%.............Figure 1.............
\begin{figure}[htbp!]
\centering
\begin{adjustbox}{max size ={\textwidth}{\textheight}}
\includegraphics[scale=0.35]{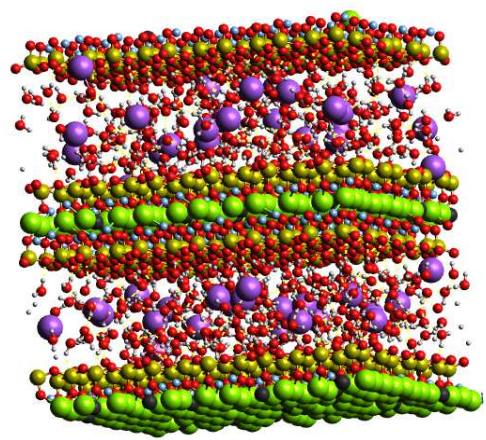}
\end{adjustbox}
\caption{Snapshot of the\ pre-equilibrated systems\ 
         with 48\ interlayer cation,\ 480 water molecule\ 
         with two half,\ and one full layer\ hectorite clay.\ 
         Colors: O (red),\ H (white),\ Si (dark yellow),\ 
         green (Mg),\ Na (purple),\ F (lightskyblue).
         \label{fig:1}}
\end{figure}
The unit-cell formula\ of Fluorohectorite clay is\  
\begin{eqnarray}
[M_{x}]^{int}[{Mg}_{6-x}{Li}_{x}]^{oct}[{Si}_8]^{tet}{O}_{20}F_{4} 
\label{eq1}
\end{eqnarray}
where $M$ denotes\ some form of\ an intercalated\ 
cation between\ clay layers.\ In the present work,\ 
the intercalated cations\ considered are monovalent\ 
cations such\ as Li$\rm^{+}$,\ Na$\rm^{+}$,\ K$\rm^{+}$,\ 
and Cs$\rm^{+}$.\ 
The simulation box\ is shown in Fig.~\ref{fig:1},\ 
and the corresponding\ crystallographic positions\ 
of the atoms\ in the clay unit\ cell are\ given\ 
elsewhere~\cite{breu2003disorder}.\ 
Certain parameters\ in real clay,\ however,\ 
remain unknown.\ These are:\ the number\ of\ 
H$\rm_{2}$O\ units per cation,\ the\ interlayer\ 
spacing,\ and\ the preferred\ orientations\ of\ 
adjacent clay\ surfaces.\ The sheets\ 
of clay\ are considered\ as rigid molecules.\ 
Our periodic\ simulation cell\ has three\ 
dimensional\ periodic boundary\ conditions.\ 
The SPC water\ model represents\ the water\ 
molecule in\ this clay-water\ interaction.\ 
The Lennard-Jones\ (L-J)\ potential model\ 
defines\ the van der Waals interactions\ 
between water-water,\ water-clay,\ water-cation,\ 
clay-clay,\ clay-cation,\ and cation-cation.\ 
Summations across\ all interaction sites\ 
yields the total\ potential energy\ for the\ 
system,\ where the\ pair-wise\ interaction\ 
is given by\
\begin{eqnarray}
V_{ij}=\frac{q_{i}q_{j}}{4\pi\varepsilon_{0} r_{ij}}
+ 4\varepsilon_{ij}\Bigg{[}
\Bigg(\frac{\sigma_{ij}}{r_{ij}}\Bigg)^{12} 
-\Bigg(\frac{\sigma_{ij}}{r_{ij}}\Bigg)^6  \Bigg{]}
\label{eq2}
\end{eqnarray}
$q_{i}$ and $q_{j}$\ are the partial charges\ 
carried by the atoms\ and ${\epsilon}_{ij}$ and\ 
${\sigma}_{ij}$ are\ the Lennard-Jones parameters\ 
obtained from\ the corresponding atomic parameters\ 
using the Lorentz Berthelot\ mixing rule~\cite{allen1989computer}.\ 
\begin{eqnarray}
{\varepsilon}_{ij}=\sqrt{{\varepsilon}_{i}{\varepsilon}_{j}} 
\label{eq3}
\end{eqnarray}
$\&$
\begin{eqnarray}
{\sigma}_{ij}=\frac{{\sigma}_{i}+{\sigma}_{j}}{2}
\label{eq4}
\end{eqnarray}
All of these\ parameters depend\ on\ 
the\ force\ field chosen\ to describe\ 
the system.\ In our study,\ we choose\ 
the flexible\ clayFF force field,\ 
slightly adapted\ to our synthetic\ 
clay,\ because it is\ shown to\ give\ 
transport\ properties of\ water,\ close\ 
to experiment.\ The\ SPC\ model for\ 
H$\rm_{2}$O\ molecules\ is\ preferred\ 
to\ be the\ TIP4P/2005 model~\cite{myshakin2013molecular}.\
The TIP4P/2005 water model\ 
has a dipole moment\ of $2.35~D$ and\ 
a polarization\ correction to the\ 
total\ energy of $5.52~{\frac{kg}{mol}}$,\ 
which is better\ as compared to\ 
other models~\cite{doi:10.1021/j100308a038}.\
When we\ simulate\ a water molecule\ inside\ 
a nanopore,\ we want to start\ out with a\ 
density close to\ the density\ of bulk water,\ 
which is~${\rho_{bulk}}= 0.99~\frac{g}{{cm}^3}$.\ 
The distribution of\ interlayer cation\ 
and water\ perpendicular to\ the plane of\ 
the clay layers\ are analysed with\ atomic\ 
density profile\ (z-density plots).\
Density profiles\ were generated to investigate\ 
the distribution\ of the various species in\ 
the clay's interlamellar space.\
The local coordination\ environment of interlayer\ 
ion by the basal\ oxygen atoms of\ the clay\ 
layer\ and water molecules\ ({H$\rm_{2}$}O)\ 
are characterized\ by RDFs.\ The radial\ 
distribution\ function for\ species\ B around\ 
species\ A is calculated\ as\ follows,\
\begin{equation}
G_{A-B}(r)=\frac{1}{4\pi \rho_{B} r^2} 
\frac{d(N_{A-B})}{dr} 
\label{eq4}
\end{equation}
where~$\rho_{B}$\ is a number density\ of species\ 
B,\ the fraction~$\frac{d(N_{A-B})}{dr}$\ is the\ 
average number\ of species particle B,\ lying in\ 
the region\ $r$ to $r+dr$,\ to\ a species particle\ 
A,\ and\ $N_{A-B}$ is\ the coordination number for\ 
species B\ around species A.\
The RDF\ gives the relative\ probability of\ 
finding\ a\ particle\ at a certain distance\ 
from a reference\ particle.\ From the\ first\ 
particle\ to first\ neighbour\ shell,\ there is\ 
no chance\ of\ penetration of atoms\ (up to\ the\ 
distance of\ diameters\ of atoms),\ so\ there\ 
is no radial\ distribution\ function and\ this\ 
region\ is\ regarded as exclusion\ region~\cite{chaikin1995principles}.\
The dynamical\ properties of\ interlayer cation\ 
and water\ is determined by\ using the mean square\ 
displacement~(MSD)\ of the ion and\ water molecule\ 
during\ the production\ run.\ The in-built in\ 
GROMACS package\ "gmx msd" is\ used to\ 
calculate the self\ diffusion coefficients.\ 
This package\ generates a data\ file of\ 
average mean\ square displacement\ as a\ 
function\ of time.\ By performing\ a linear\ 
fit of this\ data,\ we can obtain\ the slope\ 
of the straight line.\ As required by\ Einstein\ 
relation,\ dividing the slope\ by the factor\ 
6,\ see Eq.~\eqref{eq5},\ we get self diffusion\ 
coefficients\ of interlayer\ cation\ and water\ 
molecules.\    
\begin{equation}
D=\lim_{t\rightarrow\infty} {\frac{1}{6N_{m}t}}
\sum_{j=1}^{N_m} [r_{j}(t)-r_{j}(0)]^2
\label{eq5}
\end{equation}
where $N_{m}$\ is the number\ of selected\ 
species,\ $r_{j}(t)$\ is the center\ of mass\ 
position\ of the\ $j^{th}$\ species\ at time $t$.\ 
The horizontal/lateral\ diffusion coefficients\ 
were calculated using,\
\begin{equation}
D_{||} =\lim_{t{\rightarrow}{\infty}} 
\frac{\langle x^2 +y^2\rangle}{4t} 
\label{eq6}
\end{equation}
where the\ slope of\ the mean square\ 
displacement\ (MSD) parallel to\ the clay\ 
(xy plane)\ is a function\ of time between\ 
100~$ps$ and 900~$ps$.\ So far,\ it has been\ 
determined\ that the\ diffusion\ coefficient\ 
is dependent on\ temperature.\ 
The Arrhenius\ equation\ reveals\ this\ 
temperature\ dependence~\cite{Ribeiro2006BinaryMD},\
as follows,\
\begin{equation}
D = D_{0} e^{\frac{-E_{a}}{RT}}
\label{eq7}
\end{equation}
where\ $D_{0}$ is\ pre-exponential\ factor,\ 
which\ is\ also called\ frequency\ factor,\ 
$E_{a}$\ is activation\ energy for\ diffusion,\ 
and $R$ is\ universal gas constant.\ This equation\ 
can\ be further\ written as\
\begin{equation} 
E_{a} =-R \frac{\partial(\ln D)}{\partial(1/T)}
\label{eq8}
\end{equation}
The activation\ energy,\ $E_{a}$,\ can be\ 
obtained by\ a linear fit of\ $\ln(D)$ vs $\frac{1}{T}$\ 
according to Eq.~\eqref{eq8},\ by taking the\ 
slope\ of this graph\ and then multiplied\ by\ 
$-R$.\

\subsection{Molecular dynamics simulation}
Just after mixing\ water and\ ion in the\ 
layer\ of the clay,\ our system\ will be\ 
in a\ condition\ that\ is far from\ equilibrium.\ 
As a result\ of the presence\ of\ strains,\ 
it produces\ unreasonably strong\ forces\ 
between the\ atoms,\ which leads\ to\ 
the failure of\ simulation~\cite{POUDYAL201477}.\ 
The\ cause of such\ strains might be\ 
due to\ the atoms\ overlapping,\ and\ 
as a\ result,\ the\ system\ needs to\ 
be\ energy minimized.\\
When the maximum\ force on the system\ 
falls below a\ threshold value\ and the\ 
total potential\ energy is negative,\ 
the steepest descent\ procedure used\ 
to minimize energy stops.\ 
The calculations\ showing\ a negative\ 
potential\ energy are\ proof that the\ 
system is energy\ minimized.\ However,\ 
the energy reduction\ process is\ 
repeated twice\ while maintaining\ 
the same other\ circumstances and parameters\ 
as those used\ to analyse the diffusion\ 
phenomenon.\ First,\ a flexible bond is\ 
used to\ allow atoms to\ move apart from\ 
one another\ in a controlled way,\ and then\ 
a restricted bond\ is used to ensure that\ 
the new constrained\ positions do not\ 
experience strong\ forces.\ 
After energy\ minimization,\ the system\ 
is\ brought into\ temperature and\ pressure\ 
equilibrium.\ The system's thermodynamic\ 
characteristics\ change depending on\ 
many factors\ including temperature,\ 
pressure,\ density,\ etc.\ The accuracy\ 
of those\ thermodynamic properties\ to be\ 
determined would\ be impacted\ by changes\ 
to these factors.\ Consequently,\ a system\ 
must undergo\ equilibration\ before a\ 
production run~\cite{KOIRALA2022117826}.\ 
The clay-water\ system is equilibrated\ 
at four different\ temperatures ranging\ 
from 293~K to\ 350~K, and isobaric pressure\ 
of\ $1~bar$, with isothermal\ compressibility\ 
of $4.5~{\times}~10^{-5}/bar$.\ For temperature\ 
coupling,\ a velocity rescale\ thermostat\ 
is employed,\ and for pressure\ coupling,\ 
a Berendsen barostat\ is used.\ The coupling\ 
time\ constants for\ the thermostat and\ the\ 
barostat are\ $0.1~ps$ and $2.0~ps$,\ 
respectively.\ The duration\ of the equilibration\ 
is\ $1000~ps$.\\
The Particle\ Mesh Ewald (PME)\ algorithm\ 
is used for\ long range\ interactions.\ The\ 
cut-off\ parameter\ of $1.2~nm$ is\ taken with\ 
periodic boundary\ conditions for coulomb\ 
and Lennard-Jones\ (LJ) interactions.\
After\ equilibration,\ production\ run is\ 
carried\ out to\ calculate\ the thermodynamical\ 
properties\ of the\ system\ such as partial\ 
density,\ RDF,\ and diffusion\ coefficient\ 
using NVE ensemble.\
The velocity\ rescale thermostat\ is used\ 
for this run.\ All structural and dynamical\ 
quantities are\ performed for $1000~ps$ with\ 
the time step of $0.001~ps$.\
\section{Results and Discussion\label{sec:res}}
\begin{figure}[htp!]
\includegraphics[width=0.5\textwidth]{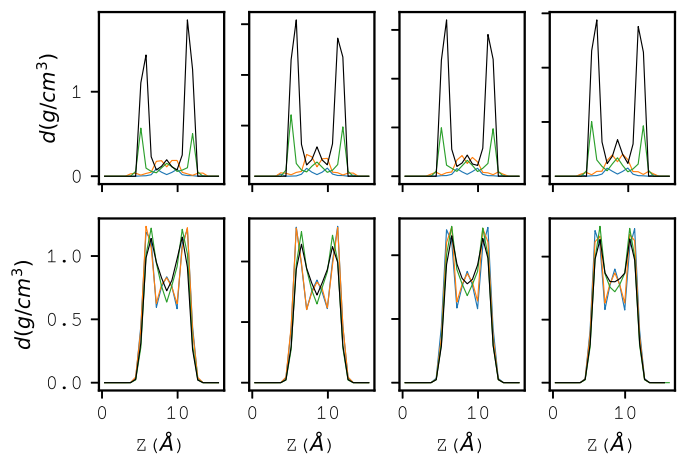}
\caption{The density profile of interlayer cation\
          water molecules\ of a Fluorohectorite\ clay\ 
          as a function\ of $z$ coordinate,\ at various\ 
          temperature\ values.\ The temperature values\ 
          start from left and the top figure\
          is for ions and below is water. \ 
          Colors:~green-KFht ,\ Orange-NaFht,\ 
          black - CsFht and blue-LiFht for\ 
          at T=350~K.\label{fig:2}} 
\end{figure}

\begin{figure*}[htbp!]
\centering
{\includegraphics[scale=0.25]{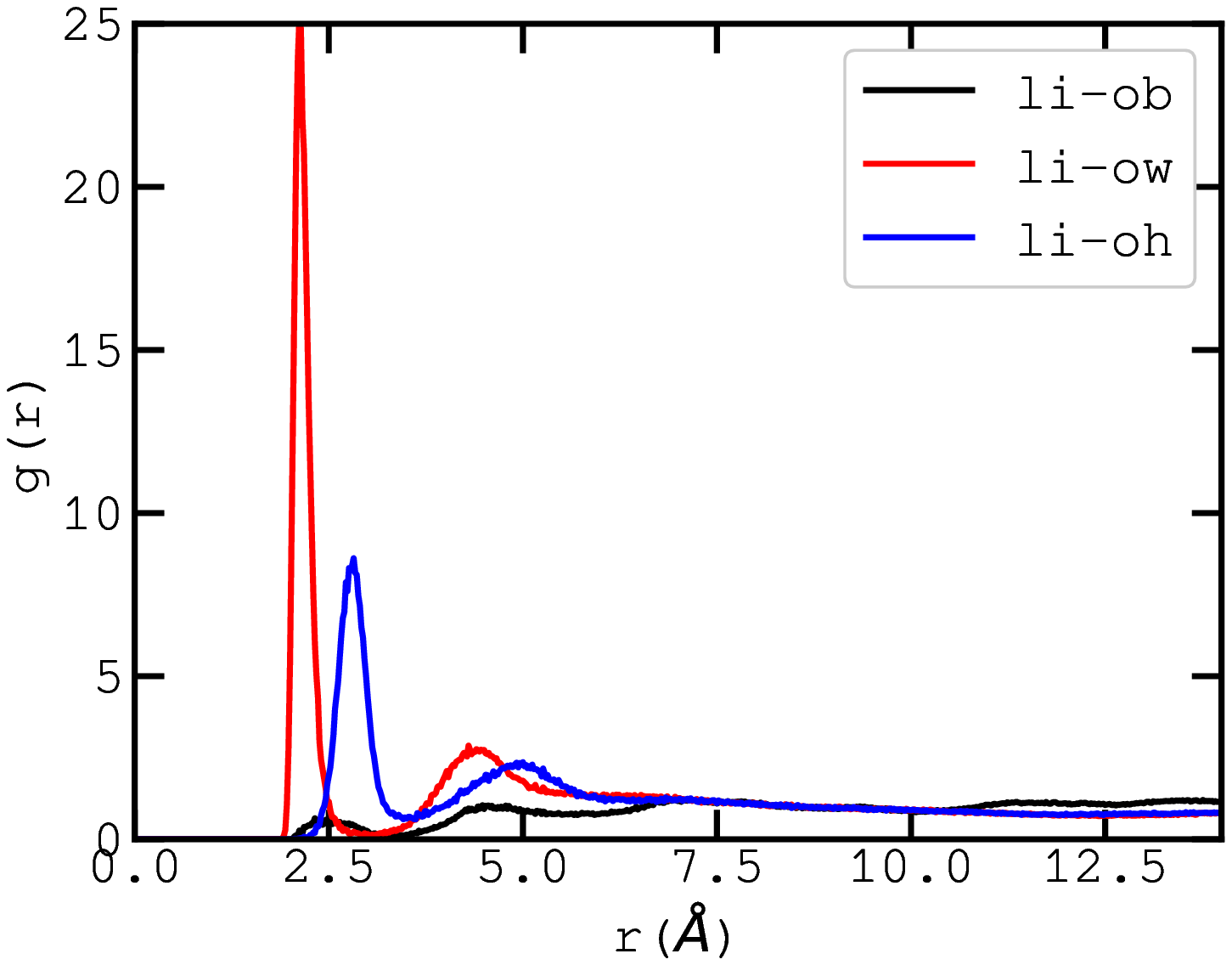}}
{\includegraphics[scale=0.25]{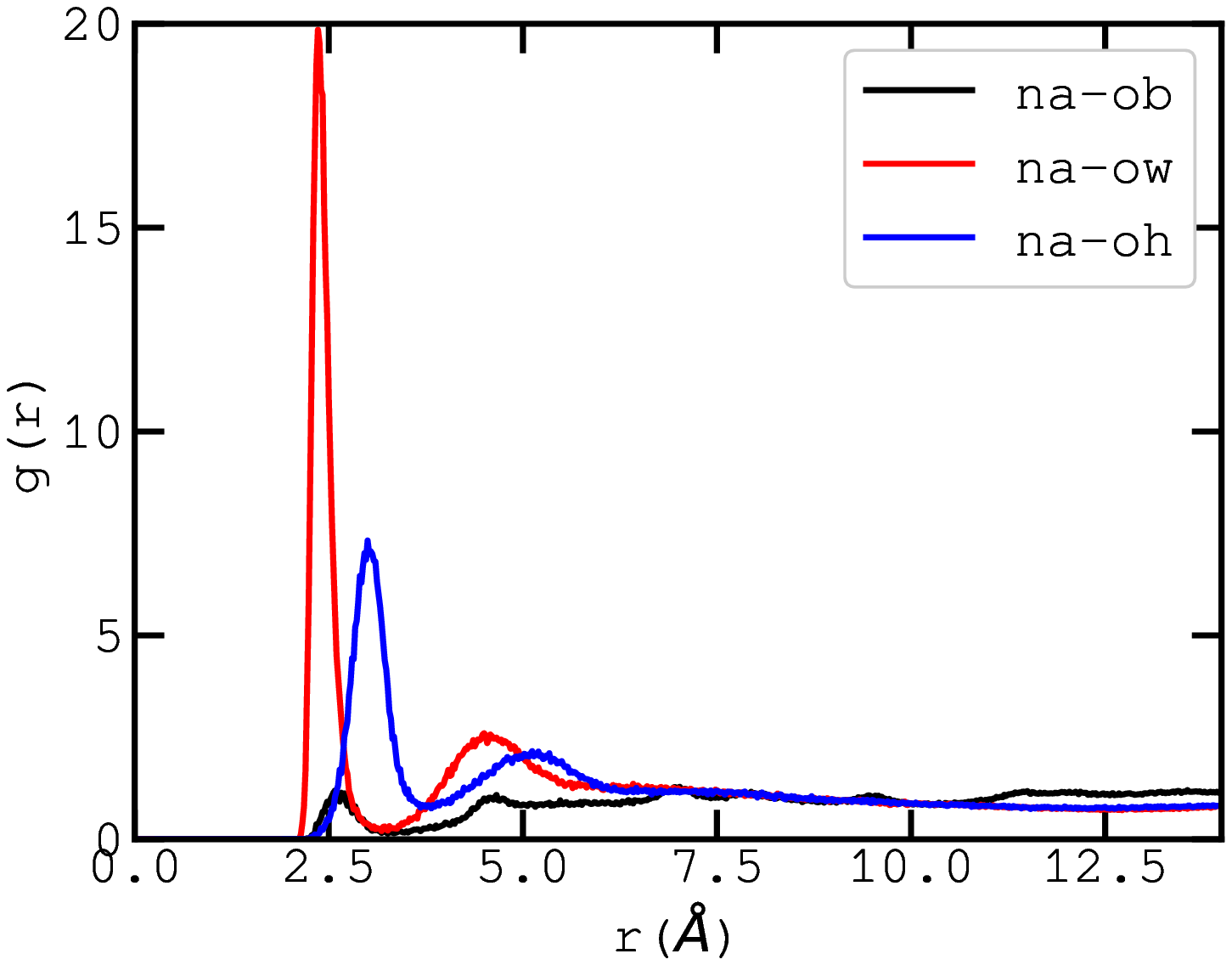}}
{\includegraphics[scale=0.25]{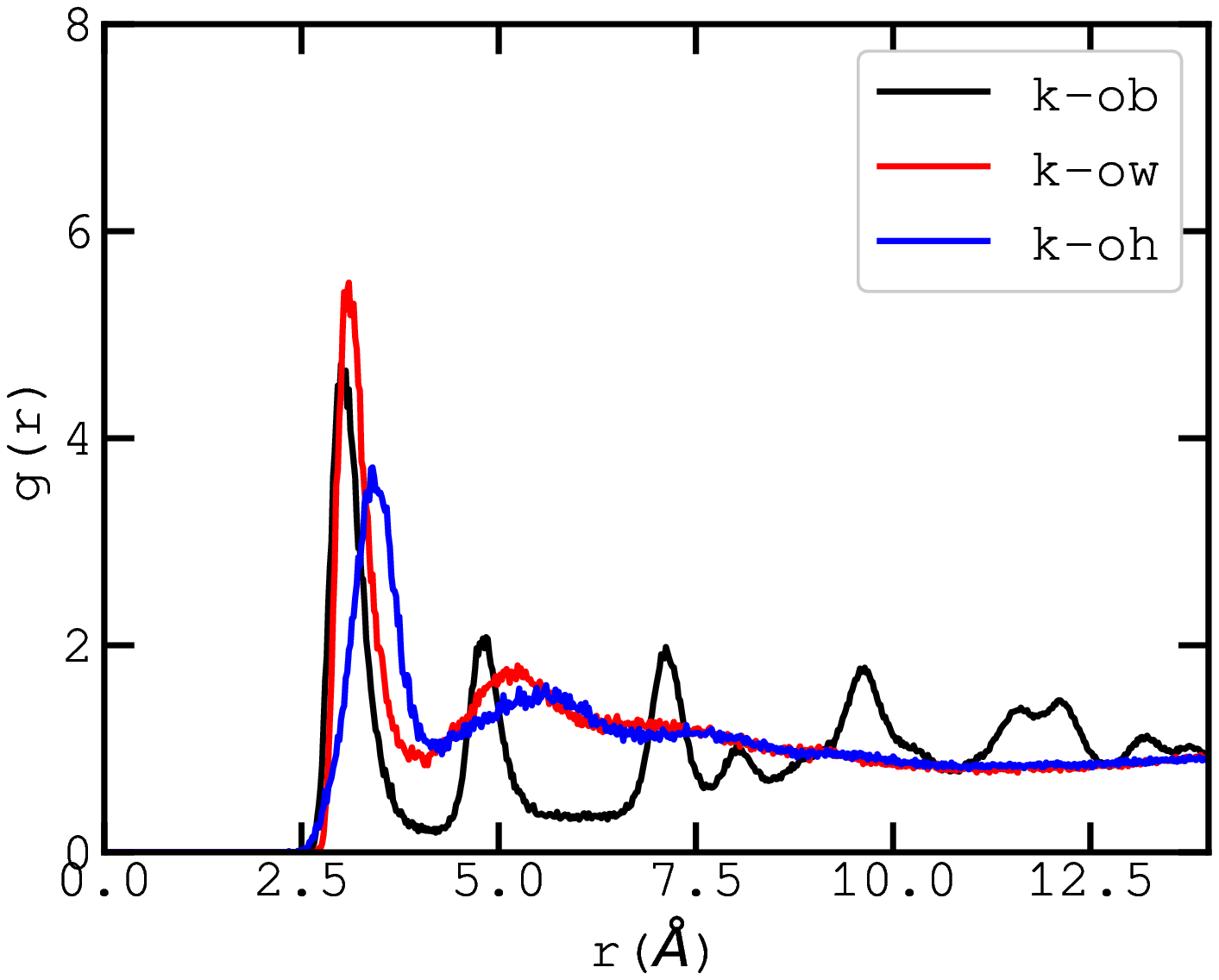}}
{\includegraphics[scale=0.25]{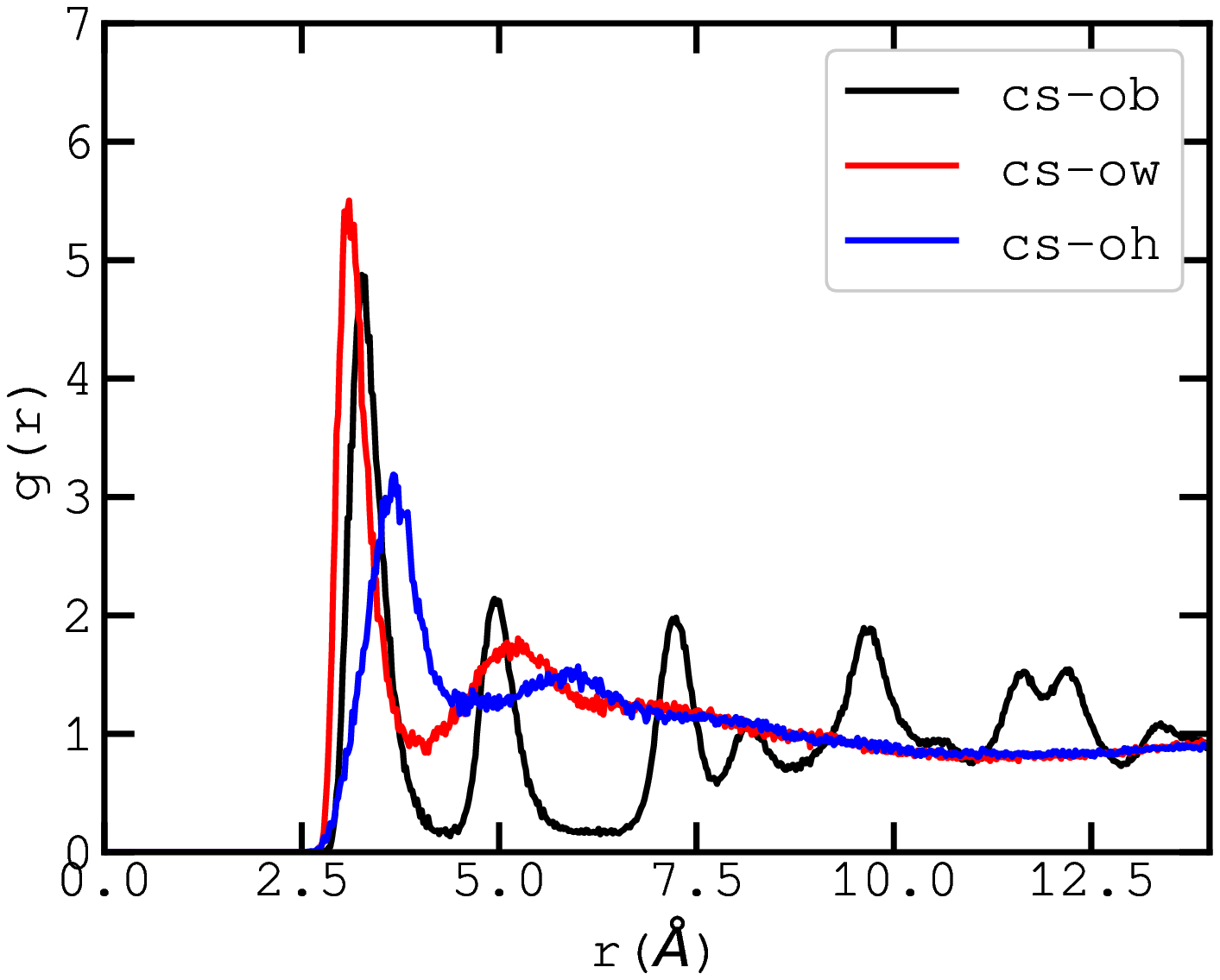}}
\caption{The RDF analysis of $g(r)$ at T=293~K.\ 
         Colors: red for cation-OW,\ black for\ 
         cation-OB,\ and blue for cation-HW.\
         Cations from left to right are 
         Li$\rm^{+}$,\ Na$\rm^{+}$,\ K$\rm^{+}$,\ 
         and Cs$\rm^{+}$,\ respectively.\ For the\ 
         interpretations\ of the\ references\ to color\ 
         in this plot legend,\ the reader is referred\ 
         to the\ web-version.\label{fig:3}}
\end{figure*}

\begin{figure*}[htbp!]
\centering
{\includegraphics[scale=0.25]{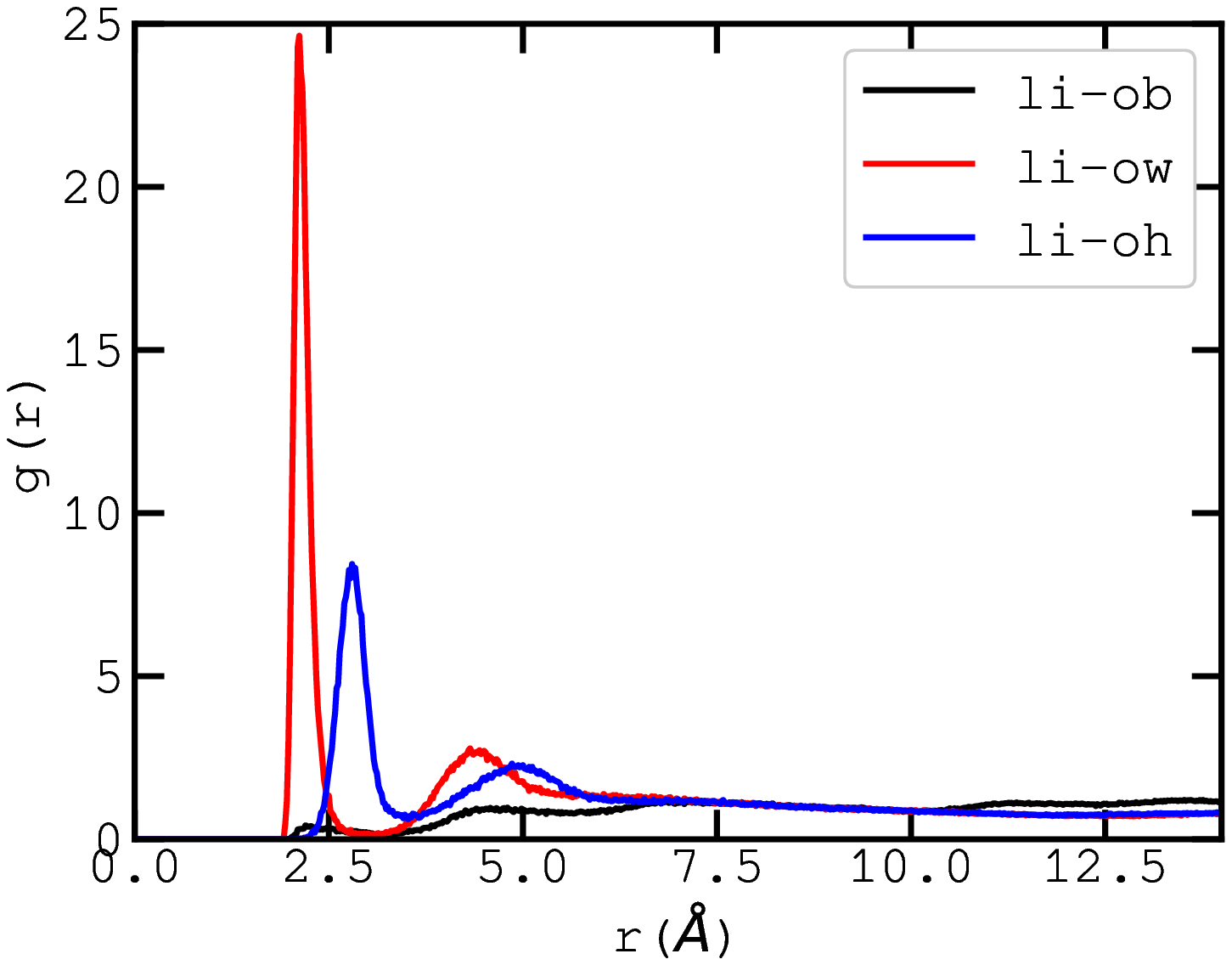}}
{\includegraphics[scale=0.25]{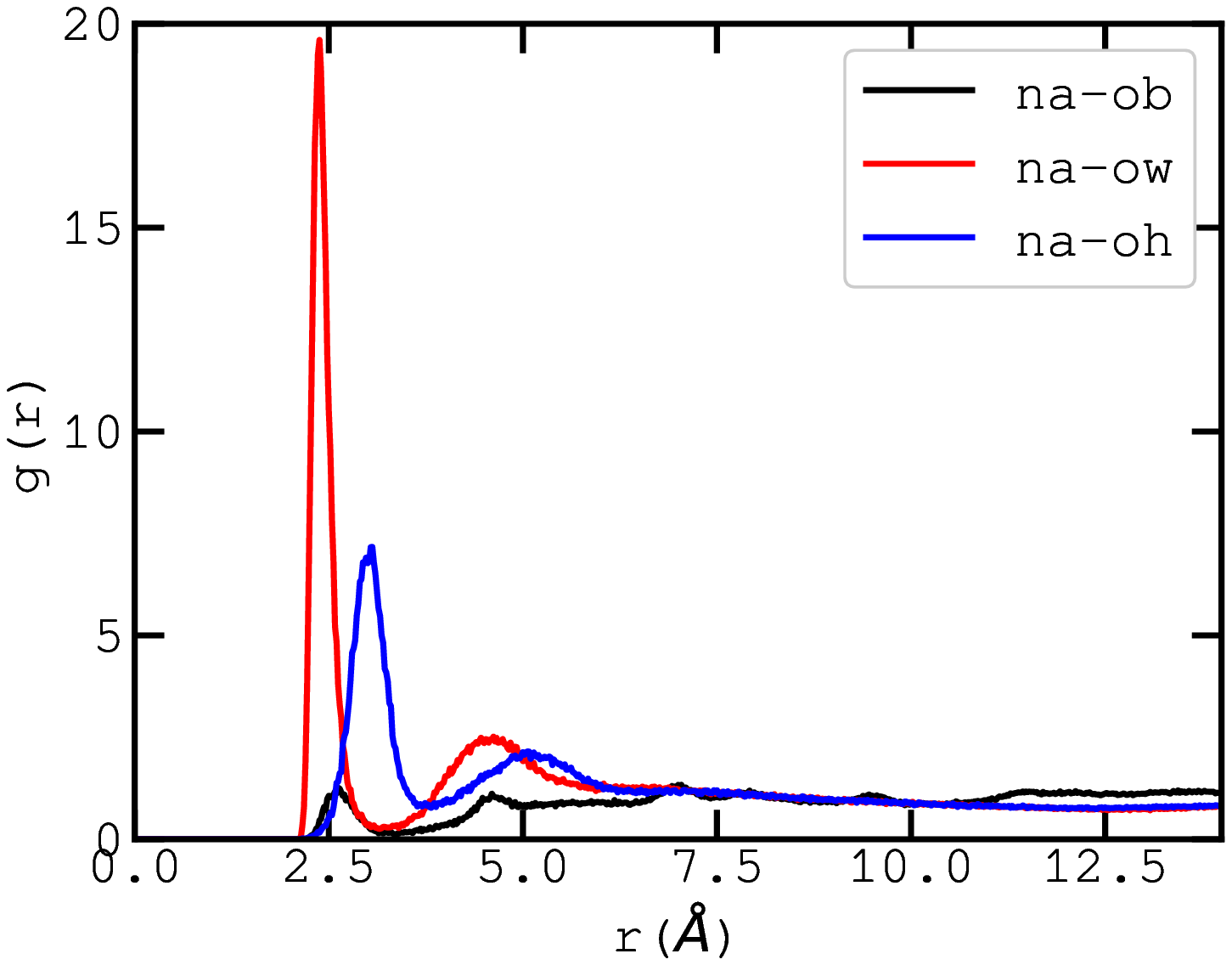}}
{\includegraphics[scale=0.25]{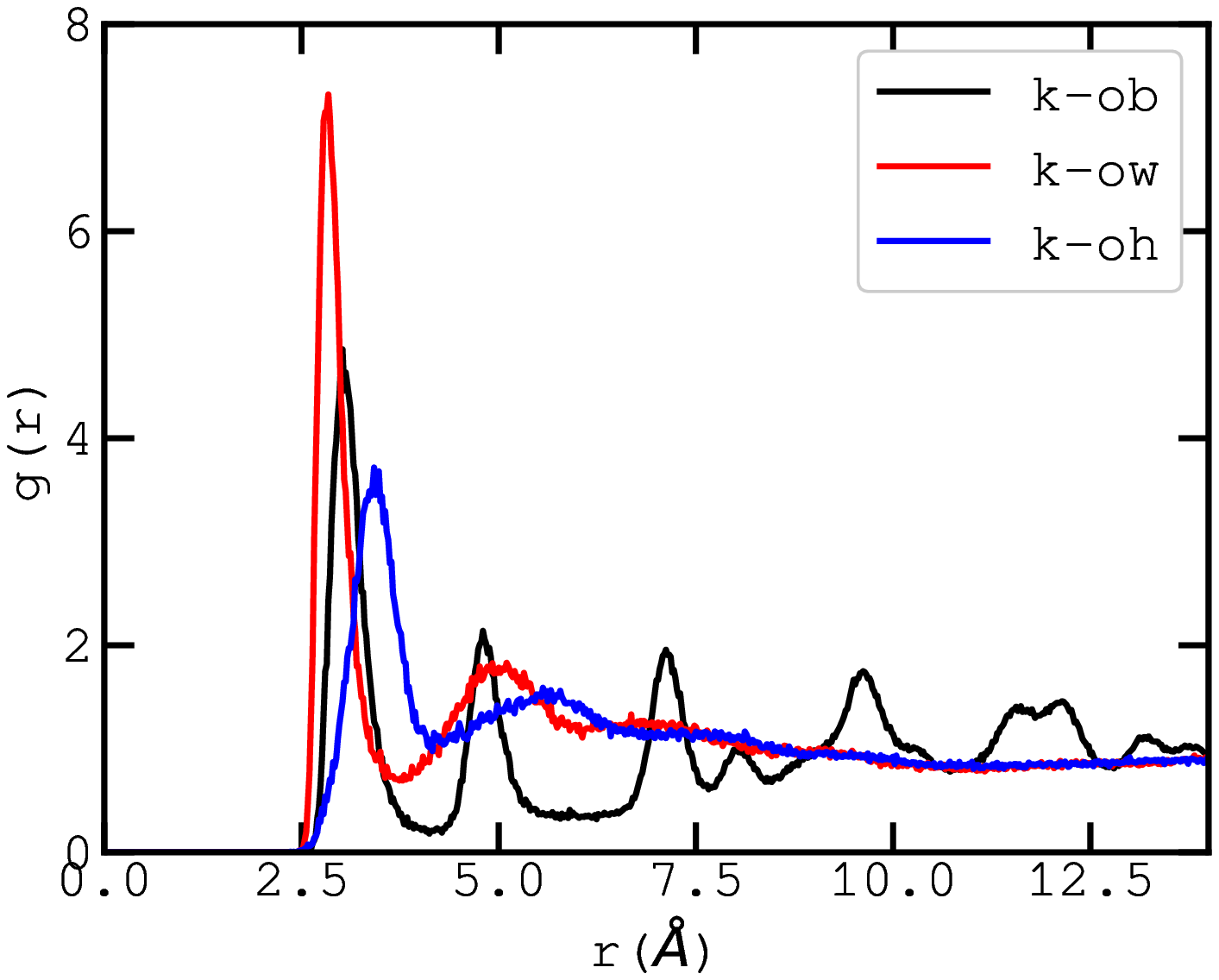}}
{\includegraphics[scale=0.25]{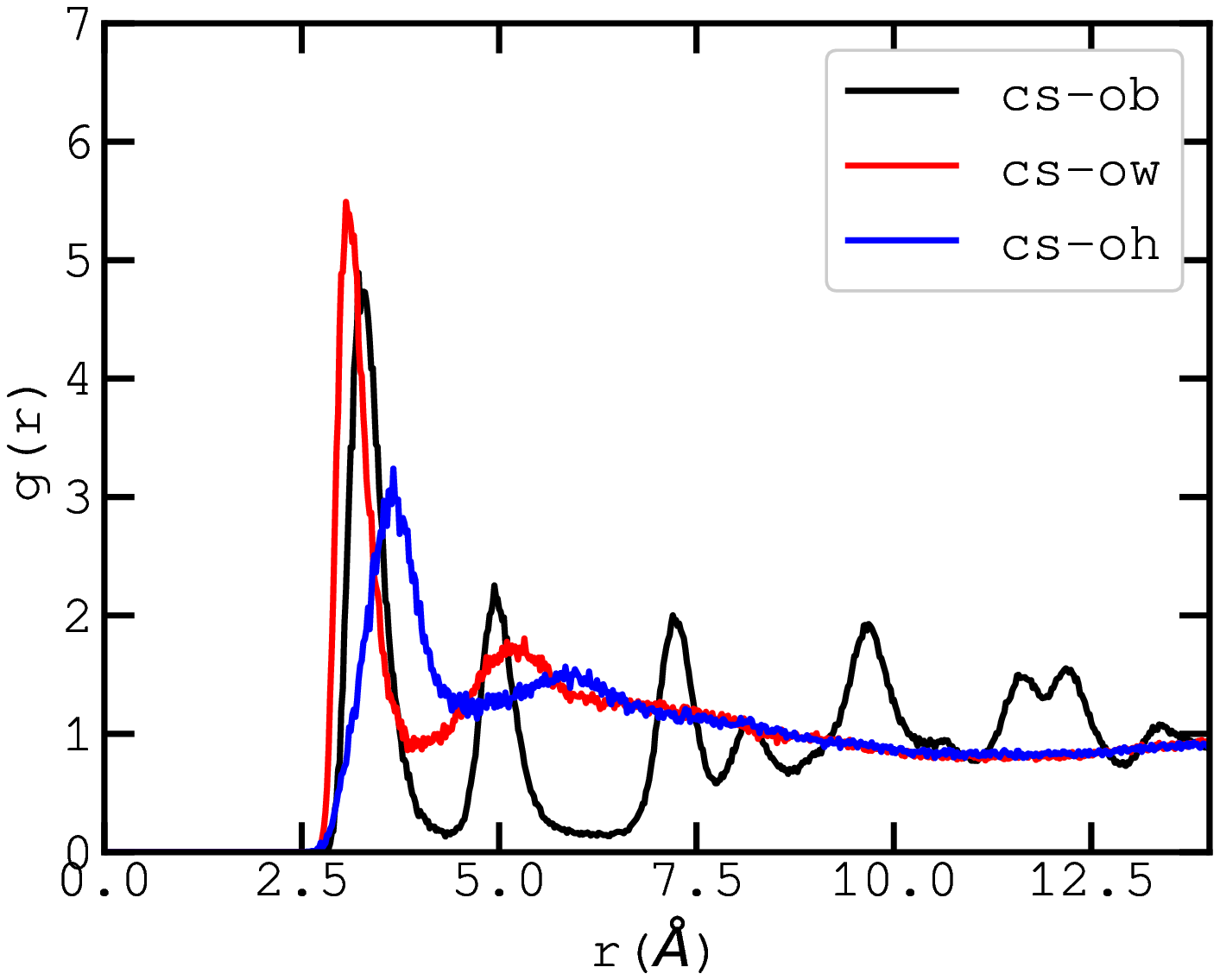}}
\caption{The RDF analysis of $g(r)$ at T=300~K.\ 
         Colors:~red for cation-OW,\ black\ 
         for cation-OB,\ and blue for cation-HW.\
         Cations from left to right are 
         Li$\rm^{+}$,\ Na$\rm^{+}$,\ K$\rm^{+}$,\ 
         and Cs$\rm^{+}$,\ respectively.\ For the\ 
         interpretations\ of the\ references\ to color\ 
         in this plot legend,\ the reader is referred\ 
         to the\ web-version.\label{fig:4}}
\end{figure*}

\begin{figure*}[htbp!]
\centering
{\includegraphics[scale=0.25]{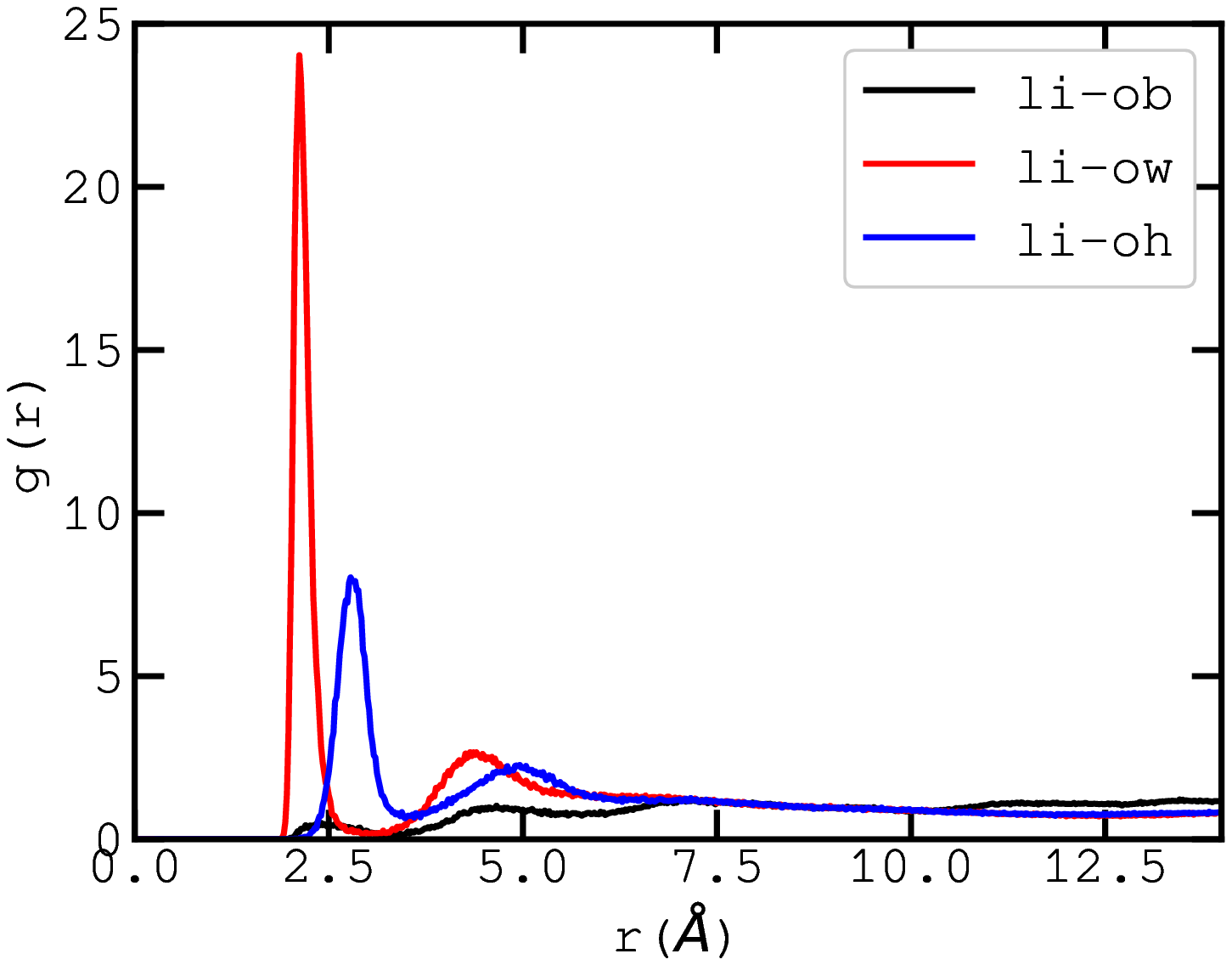}}
{\includegraphics[scale=0.25]{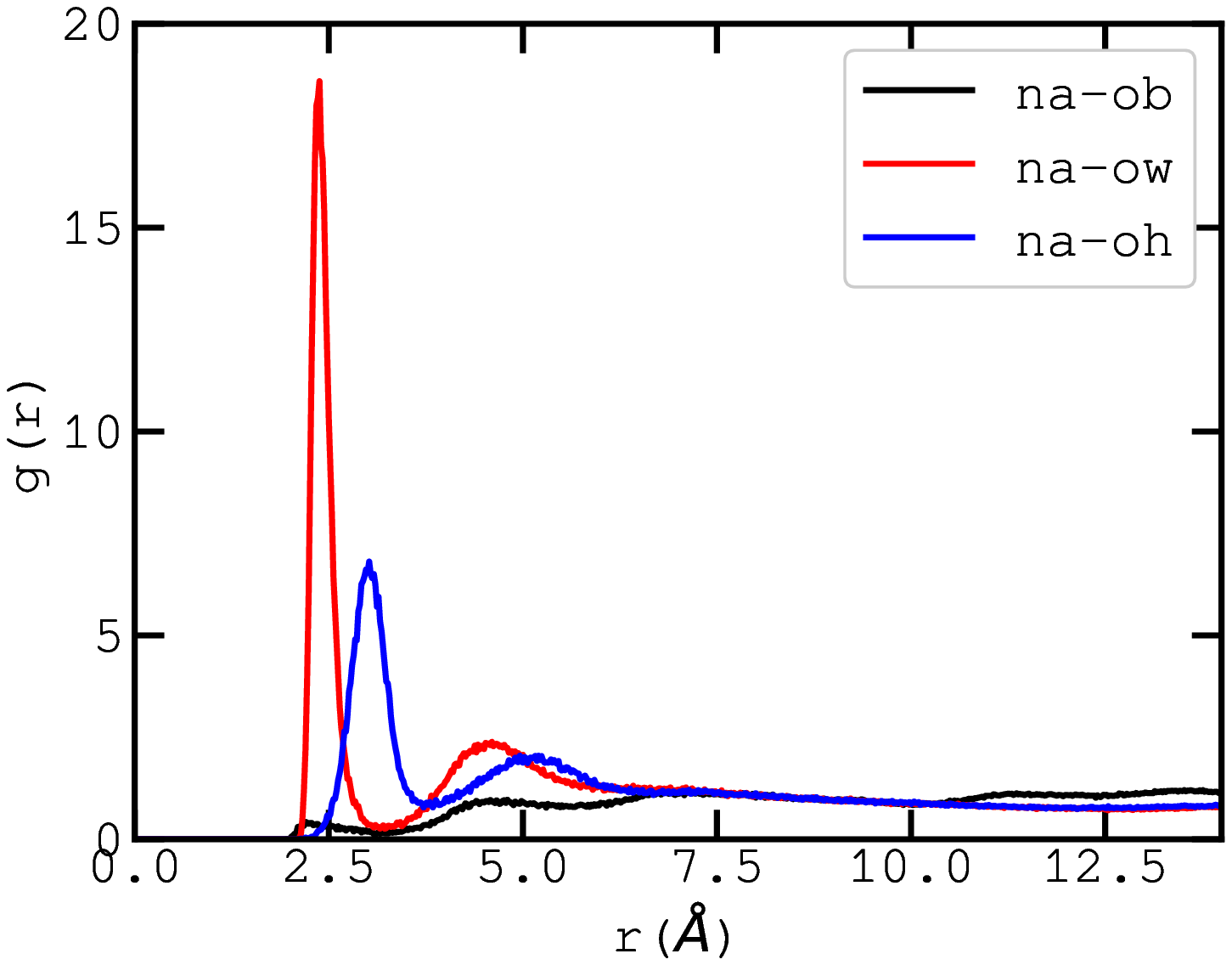}}
{\includegraphics[scale=0.25]{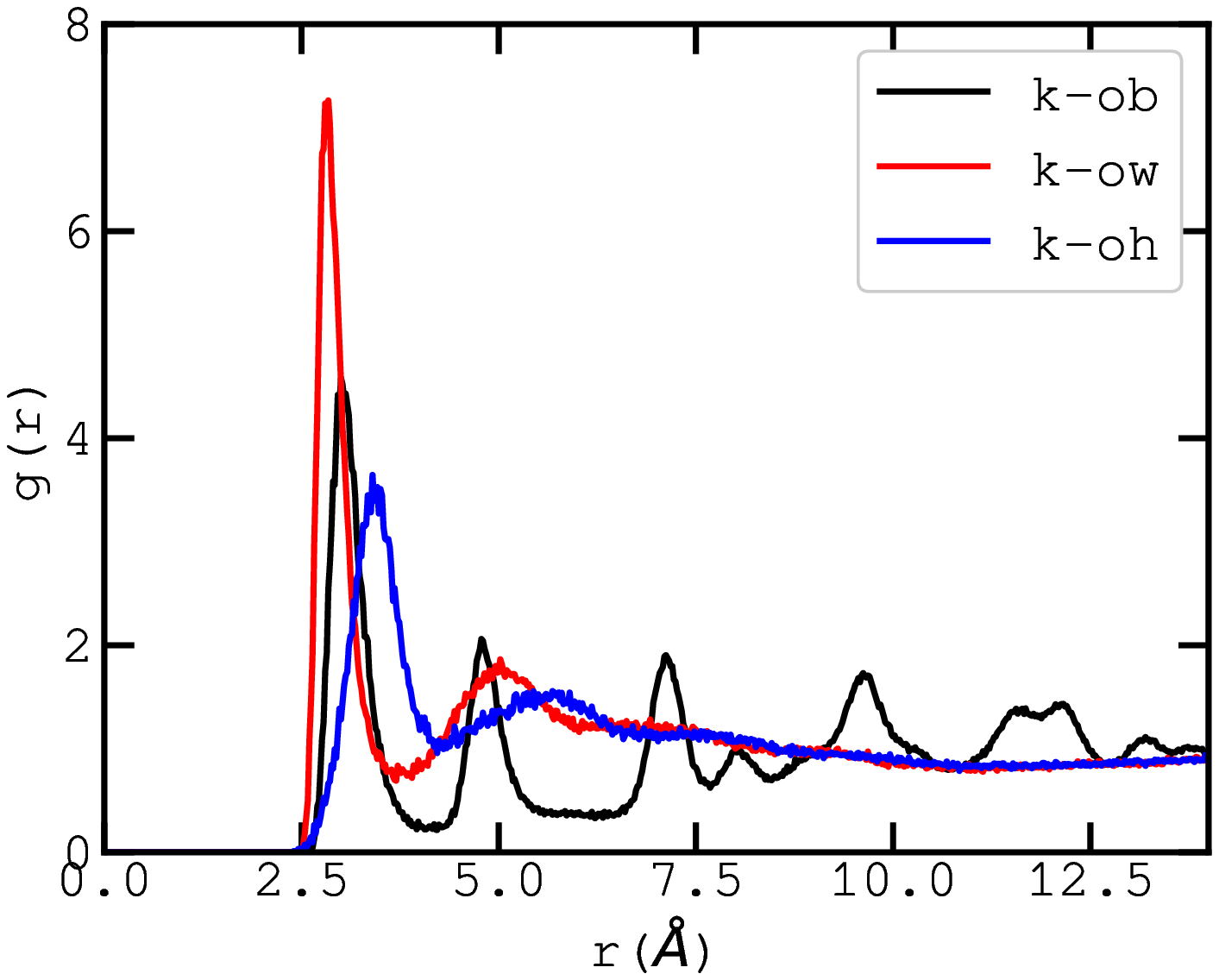}}
{\includegraphics[scale=0.25]{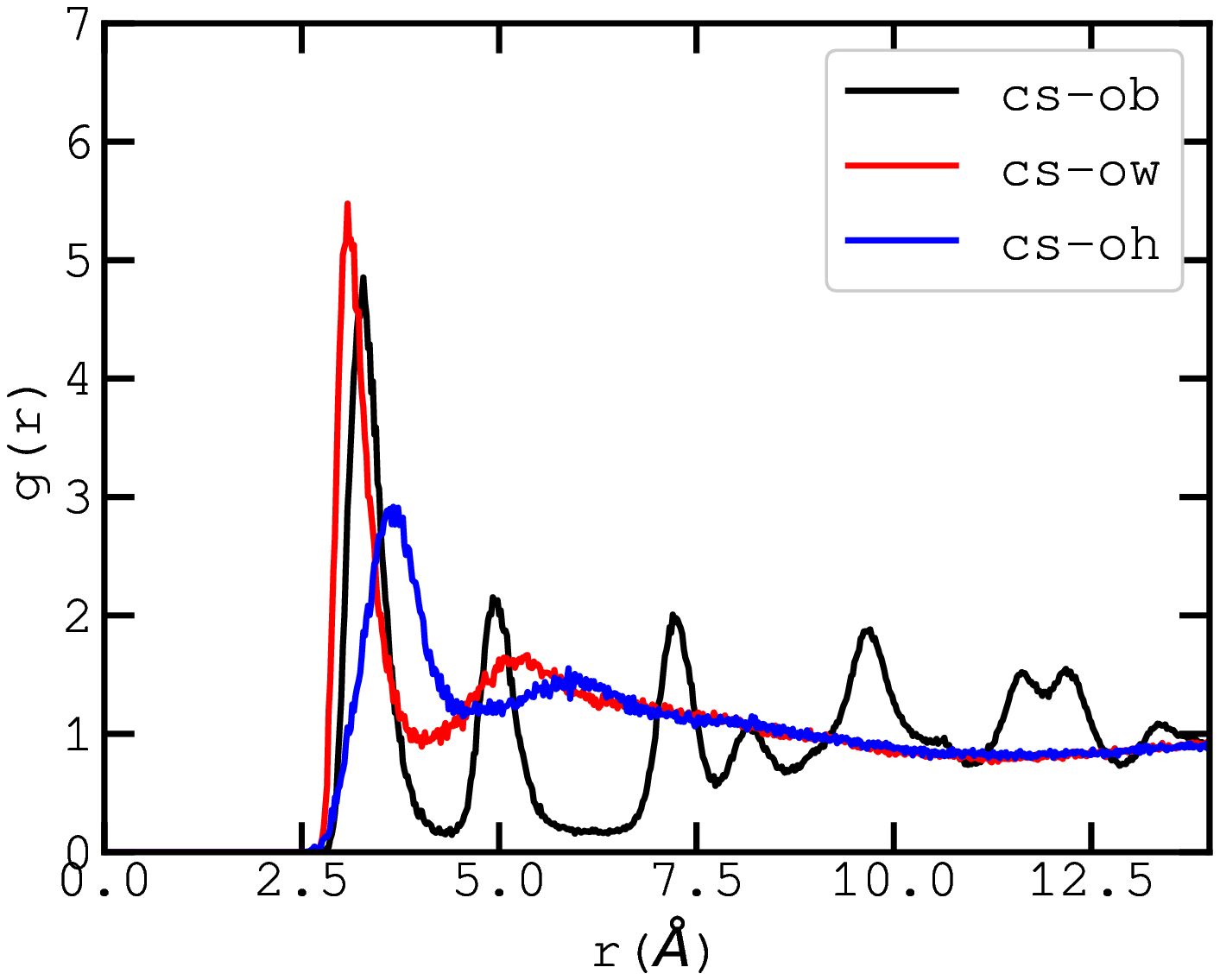}}
\caption{The RDF analysis of $g(r)$\ 
         at T=323~K.\ Colors:~red for 
         cation-OW,\ black for cation-OB,\ 
         and blue for cation-HW.\
         Cations from left to right are 
         Li$\rm^{+}$,\ Na$\rm^{+}$,\ K$\rm^{+}$,\ 
         and Cs$\rm^{+}$,\ respectively.\ For the\ 
         interpretations\ of the\ references\ to color\ 
         in this plot legend,\ the reader is referred\ 
         to the\ web-version.\label{fig:5}}
\end{figure*}

\begin{figure*}[htbp!]
\centering
{\includegraphics[scale=0.25]{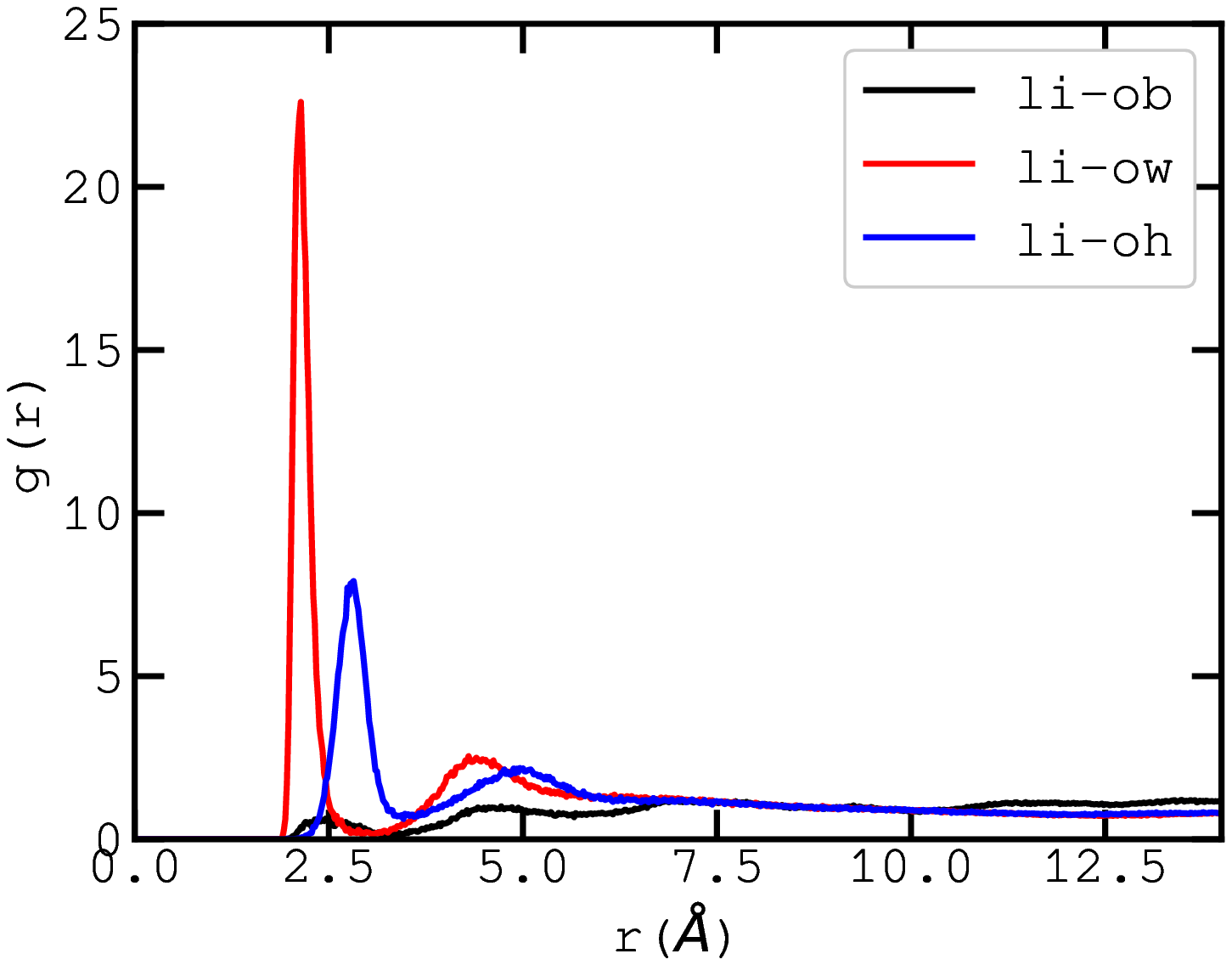}}
{\includegraphics[scale=0.25]{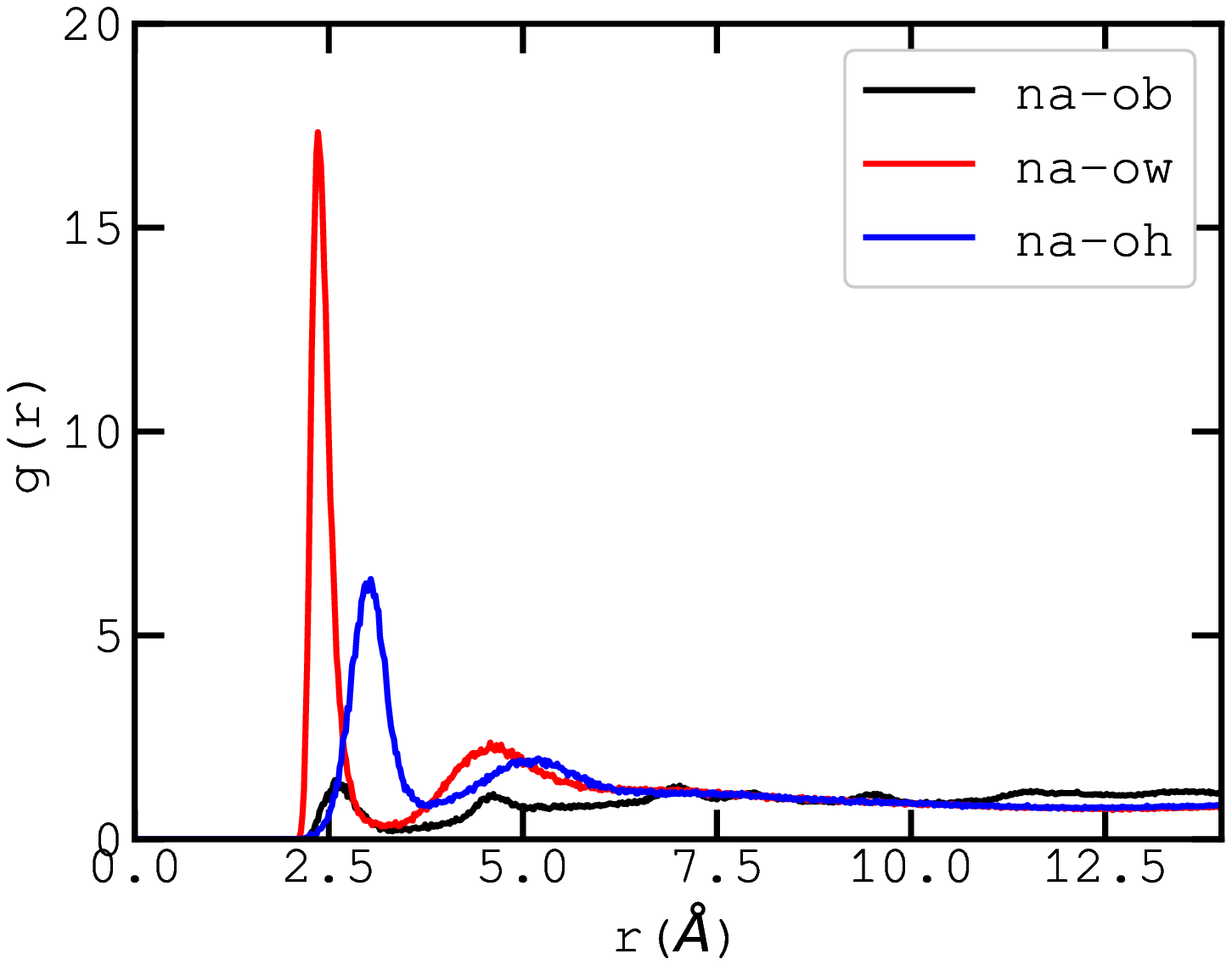}}
{\includegraphics[scale=0.25]{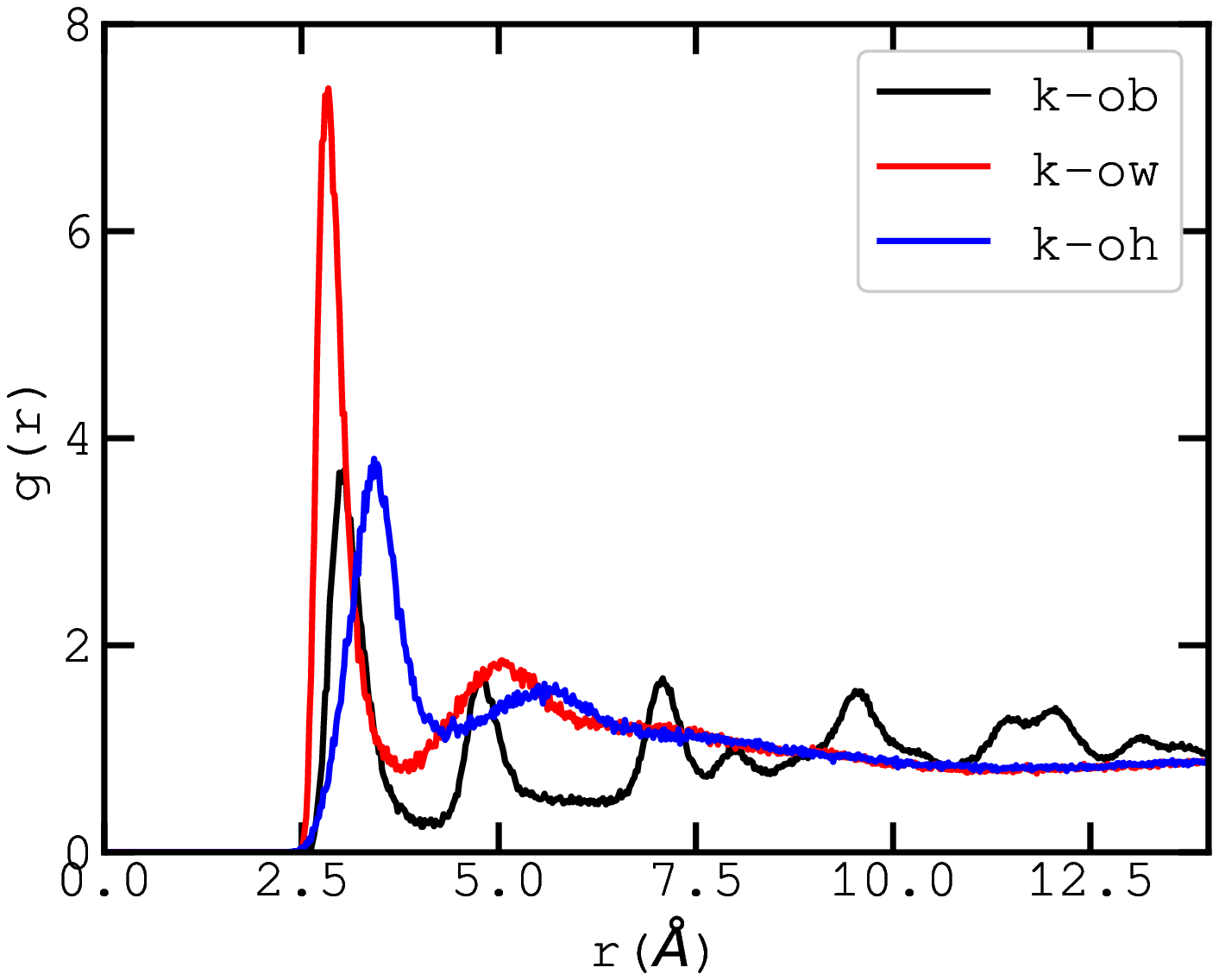}}
{\includegraphics[scale=0.25]{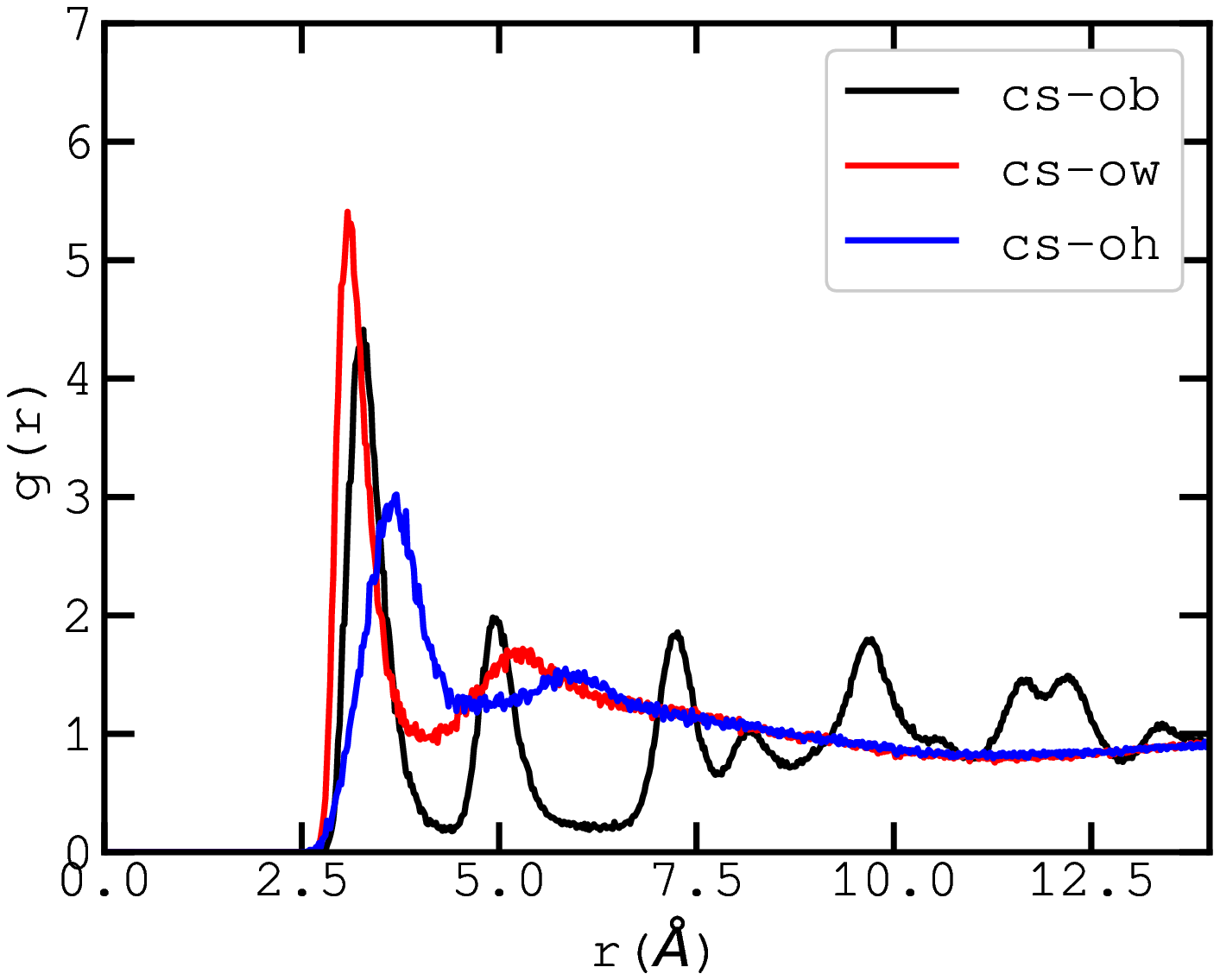}}
\caption{The RDF analysis\ of $g(r)$ at T=350~K.\ 
         Colors:~red for\ cation-OW,\ black for\ 
         cation-OB,\ and blue for cation-HW.\
         Cations from left to right are 
         Li$\rm^{+}$,\ Na$\rm^{+}$,\ K$\rm^{+}$,\ 
         and Cs$\rm^{+}$,\ respectively.\ For the\ 
         interpretations\ of the\ references\ to color\ 
         in this plot legend,\ the reader is referred\ 
         to the\ web-version.\label{fig:6}} 
\end{figure*}

Figure~\ref{fig:2}\ shows the\ calculated\ 
atomic\ density profiles of\ the interlayer\ 
ions\ and\ water at different\ temperature\ 
inside\ the\ pores\ of CsFht,\ KFht,\ LiFht,\ 
and NaFht\ clays.\ The\ profiles were\ computed\ 
along\ a direction\ perpendicular\ to the clay\ 
surface\ and averaged\ over the\ interlayer\ 
region\ of the\ simulation box\ using a\ 
1000~$ps$\ production time\ frame.\
%............Figure 2...............
A detailed analysis\ requires the radial\ 
distribution\ function~(RDF) of\ interlayer\ 
cation atoms\ with respect to\ reference\ 
oxygen atoms\ of water,\ which  are shown\ 
in Figs.~\ref{fig:3}-\ref{fig:6},\ with\ 
red color.\ For K and\ Cs-Fluorohectorite,\ 
the 1$\rm^{st}$\ neighbouring peak\ has\ 
$g(r)$ peak intensity values\ of\ 7.64\ 
at T=293~K;\ 7.30~at T=300~K;\ 7.22~at T=323~K;\ 
7.33~at T=350~K;\ 5.47~at T=300~K;\ 
5,46~at T=300~K;\ 5.45~at T=323~K;\ 
and 5.38~at T=350~K.\ However,\ 
for Li and\ Na-Fluorohectorite,\ the\ 
1$\rm^{st}$\ neighbouring\ peak\ has\ 
$g(r)$\ peak\ intensity values\ of\ 
25.13~at T=293~K;\ 24.50~at T=300~K;\ 
23.92~at T=323~K;\ 22.45~at T=350~K;\  
19.76~at T=293~K,\ 19.55~at\ T=300~K;\ 
18.47~at T=323~K;\ and 17.24~at T=350~K.\ 
This result\ simply shows\ that the peak\ 
intensity\ change\ owing\ to the presence\ 
of different\ interlayer cations.\ In addition\ 
to this,\ comparing\ with the hydration\ 
shell around K$\rm^{+1}$ and Cs$\rm^{+1}$\ 
cations,\ we found that\ the $1^{st}$ neighbour\ 
cation oxygen\ are assembled more\ densely than\ 
compared to those in the case of\ Li$\rm^{+1}$\ 
and Na$\rm^{+1}$\ cations,\ which is the reason\ 
why the\ $1^{st}$\ neighbour\ peak of K and\ 
Cs-Fluorohectorite\ are less\ than that of\ 
Li and Na-fluorohectorite.\
It is well\ known that\ the interaction of\ 
the clay\ cation with their\ environment\ 
depends on\ non-bonding\ electrostatics\ 
and\ van der Waals potentials.\ 
As shown in\ the Figures\ discussed above,\ 
the RDFs between the\ cation and the clay\ 
tetrahedral oxygen\ atom of the hydrated\ 
clay\ clearly show\ the structure.\ Similarly,\ 
the RDF\ of interlayer\ cation with reference\ 
to basal\ oxygen of\ clay is\ also plotted\ 
and\ represented by\ black color in all\ 
the Figs.~\ref{fig:3}-\ref{fig:6}.\ 
We can see\ the exclusion\ region up to a\ 
certain\ distance,\ where the probability\ 
of finding\ the particle with\ respect to\ 
the\ reference particle\ is zero.\ As we go\ 
to\ right\ side of each point,\ the height\ 
of\ the peaks\ keep on decreasing.\ This\ 
means\ that the\ distance between\ the\ 
reference\ atom and\ atom\ under the study\ 
increases.\ The correlation\ between them\ 
decreases\ and\ eventually there\ won't be\ 
any\ long range\ correlation for\ a large\ 
value of\ $r$,\ i.e.,\ $g(r)~{\rightarrow}~1$~\cite{chandler1987introduction}.\
The exclusion region\ for cation-OW and\ 
cation-OB correlation\ pair are provided\ 
in Table~\ref{tab1}\ at all temperature.\
%..............Table 1...................    
\begin{table*}[!htbp]
\addtolength{\tabcolsep}{0.001mm}
\renewcommand{\arraystretch}{1.2}
\centering 
\caption{The exclusion region of each correlation pair 
         in {\AA} for different temperatures.
         \label{tab1}}
\begin{tabular}{|c|c|c|c|c|}
\hline
\multirow{2}*{Correlation pair} & \multicolumn{4}{c|}{Temperature~[K]}\\
\cline{2-5}
                 &  293 K & 300 K & 323 K & 350 K\\ 
\hline
Cs-OW            & 0.267 & 0.269 & 0.272 & 0.268 \\ 
\hline
Cs-OB            & 0.278 & 0.278 & 0.279 & 0.278 \\ 
\hline
K-OW             & 0.249 & 0.250 & 0.250 & 0.250 \\ 
\hline
K-OB             & 0.259 & 0.260 & 0.260 & 0.260 \\ 
\hline
Li-OW            & 0.189 & 0.191 & 0.197 & 0.196 \\ 
\hline
Li-OB            & 0.206 & 0.197 & 0.197 & 0.196 \\ 
\hline
Na-OW            & 0.214 & 0.215 & 0.213 & 0.209 \\ 
\hline
Na-OB            & 0.216 & 0.220 & 0.220 & 0.216 \\
\hline
\end{tabular}
\end{table*}
The coordination\ number of interlayer cations\ 
is the number of molecule\ of water and basal\ 
oxygen of the\ clay in the immediate neighbour\ 
of the cations.\ It depends on the distance\ 
between the water\ molecule as well as the\ 
basal oxygen\ of clay and cations.\
In Tables~\ref{tab2}~$\&$~\ref{tab3},\ 
the coordination number\ of each correlation\ 
pair is presented.\
%..............Table 2...................
\begin{table*}[!htbp]
\addtolength{\tabcolsep}{2.2mm}
\renewcommand{\arraystretch}{1.4}
\centering 
\caption{The coordination number of different clays with 
         different temperature.\label{tab2}} 
\begin{tabular}{|l|l|l|l|l|l|l|}
\hline
\multirow{2}*{T~[K]} & \multicolumn{3}{c|}{LiFht} & \multicolumn{3}{c|}{NaFht}\\ 
                      \cline{2-7} 
                      & {Li-OW} & {Li-OB} & {Li-HW} & {Na-OW} & {Na-OB} & {Na-HW}\\ 
\hline
293                   & {1.93}  & {1.23}  & {1.67}  & {1.85} & {1.24}  & {1.65}\\ 
\hline
300                   & {1.90}  & {1.22}  & {1.64}  & {1.85} & {1.24}  & {1.64}\\ 
\hline
323                   & {1.92}  & {1.23}  & {1.67}  & {1.83} & {1.24}  & {1.63}\\ 
\hline
350                   & {1.90}  & {1.24}  & {1.65}  & {1.79} & {1.25}  & {1.60}\\ 
\hline
\end{tabular}
\end{table*}
%..............Table 3...................
\begin{table*}[!htbp]
\addtolength{\tabcolsep}{2.2mm}
\renewcommand{\arraystretch}{1.4}
\centering 
\caption{The coordination number of different clays with 
         different temperature.\label{tab3}} 
\begin{tabular}{|l|l|l|l|l|l|l|}
\hline
\multirow{2}*{T~[K]} & \multicolumn{3}{c|}{KFht} & \multicolumn{3}{c|}{CsFht}\\ 
\cline{2-7} 
                     & {K-OW} & {K-OB} & {K-HW} & {Cs-OW} & {Cs-OB} & Cs-HW\\ 
\hline
293                  & {1.54} & {1.24} & {1.44} & {1.49} & {1.32} & {1.42}\\ 
\hline
300                  & {1.54} & {1.34} & {1.45} & {1.49} & {1.32} & {1.42}\\ 
\hline
323                  & {1.54} & {1.34} & {1.45} & {1.49} & {1.32} & {1.42}\\ 
\hline
350                  & {1.55} & {1.30} & {1.46} & {1.50} & {1.32} & {1.42}\\ 
\hline
\end{tabular}
\end{table*}
%.............Figure 3.................
%.............Figure 4.................
%.............Figure 5..................
%.............Figure 6.................
  
%.............Figure 7.................
\begin{figure*}[!htbp]
\centering
{\includegraphics[scale=0.4]{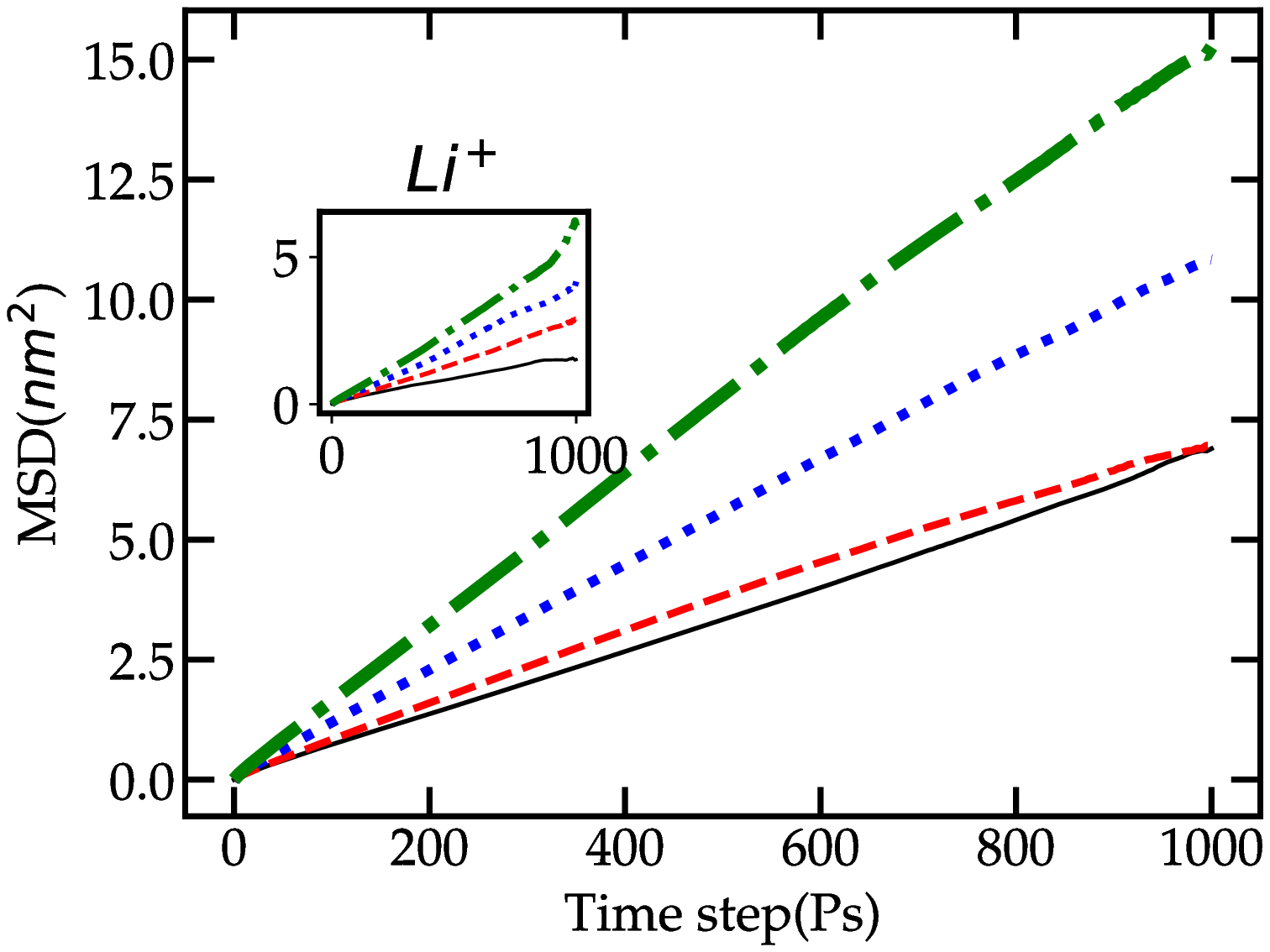}} 
{\includegraphics[scale=0.4]{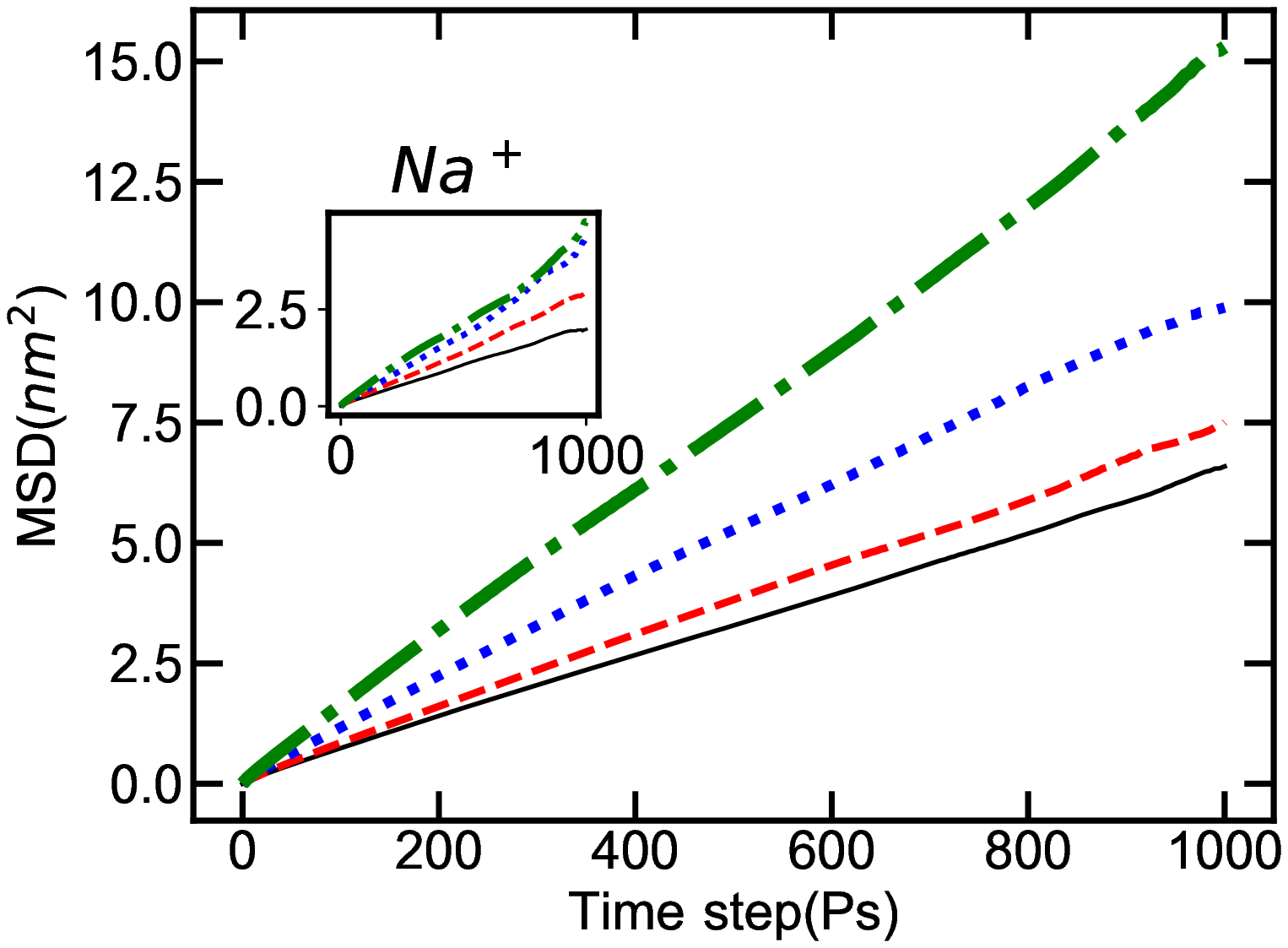}} 
{\includegraphics[scale=0.4]{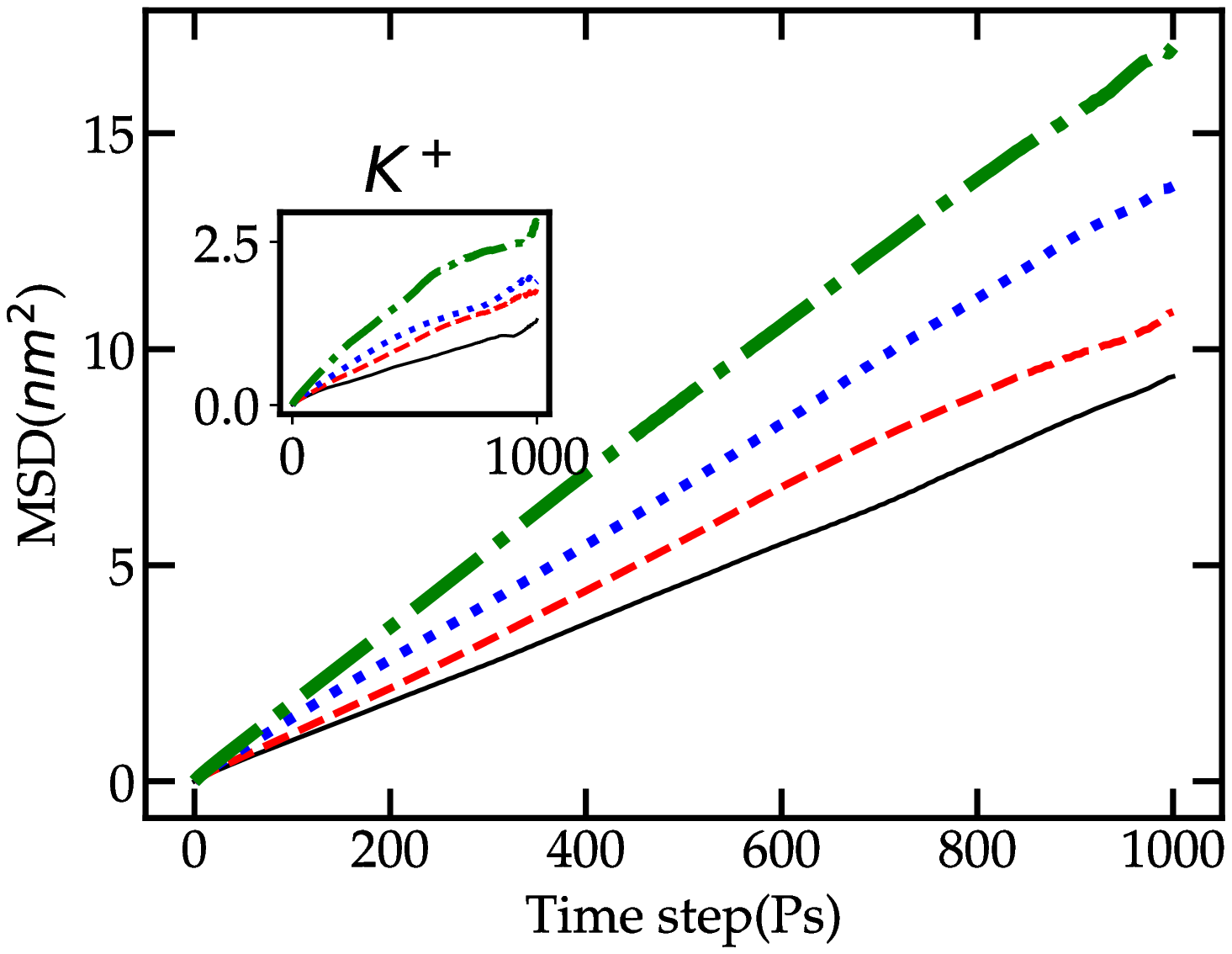}} 
{\includegraphics[scale=0.4]{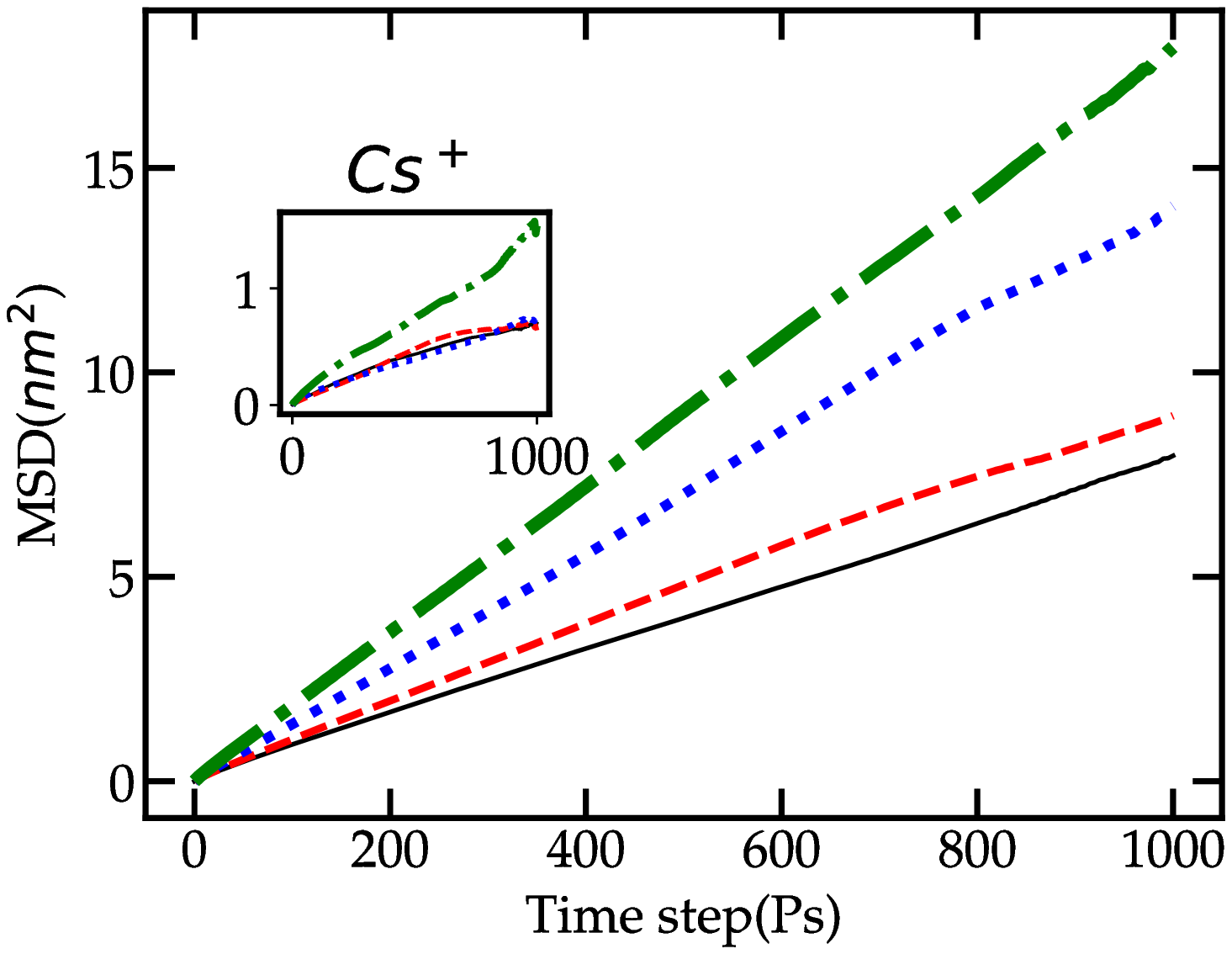}} 
\caption{The mean square\ displacement graph as a\ 
         function of time step~($ps$)\ of interlayer\ 
         cations~$\&$~water.~Colors:~black for T=393~K,\ \ 
         red for T=300~K,\ blue for T=323~K and green for\ 
         T=350~K.\label{fig:7}}   
\end{figure*}
The Figure~\ref{fig:7} show\ 
the mean\ square displacement\ (MSD) plots of\ 
interlayer cations\ (Cs$\rm^{+}$,\ K$\rm^{+}$,\ 
Li$\rm^{+}$,\ Na$\rm^{+}$),\ and H$\rm_{2}O$\ 
molecule in\ CsFht,\ KFht,\ LiFht and NaFht,\ 
at different\ temperatures ranging from 293~K\ 
to 350~K,\ respectively.\ The corresponding\ 
simulated result\ of self diffusion\ coefficients\ 
in 3D and 2D\ are shown in\ Tables~\ref{tab4}~$\&$~\ref{tab5},\ 
respectively.\ It is clearly\ observed that\ 
when\ temperature increases,\ the generated\ 
velocities\ of the ions\ and water molecules\ 
also increase\ and the density\ of the\ system\ 
decreases.\ This\ provides more\ space for the\ 
molecule to execute\ random walk.\ Due to this,\ 
the mean square\ displacement increases.\ From\ 
Einstein's\ relation in Eq.~\eqref{eq5},\ as the\ 
mean\ square displacement increases,\ the self\ 
diffusion\ coefficient also increases.\\
Comparing the\ simulated results\ of the diffusion\ 
coefficients of water\ with the experiment and\ 
other simulation values,\ at T=300~K,\ water molecule\ 
within the bilayer\ has self diffusion coefficient\ 
of $[1.0-1.5]~{\times}~10^{-5} {cm^{2}}s^{-1}$.\ 
According\ to\ a literature~\cite{suter2015ab},\ 
it is varies\ from\ $[1.96-2.33]~{\times}~10^{-6} {cm^{2}}s^{-1}$.\ 
Water\ diffusion coefficients\ in a\ synthetic\ 
hectorite\ clay\ is investigated by a\ 
literature~\cite{malikova2005diffusion},\ 
to be\ $[4.6~{\pm}~0.3]~{\times}~10^{-6} {cm^{2}}s^{-1}$~(NSE)\ 
and\ $[4.3~{\pm}~0.1]~{\times}~10^{-6} {cm^{2}}s^{-1}$~(TOF).\ 
But\ according to the\ recent measurement of (TOF)\ 
on\ bihydrated vermiculite,\ the results are\ 
slightly\ deviated from\ the above values,\ i.e.,\ 
it is\ $2.7~{\times}~10^{-5} {cm^{2}}s^{-1}$~\cite{swenson2000quasielastic}.\ 
The diffusion coefficient\ of water in sodium\ 
vermicutite~(Na-Ver)\ and cesium vermicutite~(Cs-Ver)\ 
is $1.6~{\times}~10^{-6} {cm^{2}}s^{-1}$ and\ 
$1.8~{\times}~10^{-6} {cm^{2}}s^{-1}$,\ 
respectively.\ From bihydrated (Na$\rm^{+}$,\ Cs$\rm^{+}$)\ 
montmorilonite clay,\ the result is $[1.2-1.5]~{\times}~10^{-5}{cm^2}s^{-1}$~\cite{marry2003microscopic,chang1995computer}.\ 
Our result of\ diffusion\ coefficient of water\ 
in Na-Fht and Cs-Fht at 300~K\ is\ 
$[1.20~{\pm}~0.06]~{\times}~10^{-5} {cm^{2}}s^{-1}$\ 
and $[1.52~{\pm}~0.18]~{\times}~10^{-5} {cm^{2}}s^{-1}$,\ 
respectively.\\
The actual diffusion\ coefficient of water\ is\ 
$2.27~{\times}~10^{-5}{cm^2}s^{-1}$.\ 
These values of\ diffusion coefficients\ are\ 
higher than half\ of the bulk value\ for water,\ 
which is $D_{exp}^{bulk-{H_{2}}O}=2.3~{\times}~10^{-5} 
{cm^{2}}s^{-1}$.\
This is true\ for three types of clay studied\ 
(hectorite, montmorillonite, vermiculite) and\ 
both mono and divalent counterions~\cite{tuck1985quasi,
zheng2011water,malikova2005diffusion}.\
In Table~\ref{tab4},\ the diffusion coefficient\ 
of interlayer cation\ of synthetic fluorohectorite\ 
clay is presented.\ From this,\ one can see that the\ 
diffusivity of\ Na$\rm^{+}$ $>$ Li$\rm^{+}$ $>$\ 
K$\rm^{+}$ $>$ Cs$\rm^{+}$\ at T=293K,\ Na$\rm^{+}$ $>$\ 
Li$\rm^{+}$ $>$ K$\rm^{+}$ $>$ Cs$\rm^{+}$\ at T=300~K,\ 
Li$\rm^{+}$ $>$ Na$\rm^{+}$ $>$ K$\rm^{+}$ $>$ Cs$\rm^{+}$\ 
at T=323~K,\ and Li$\rm^{+}$ $>$ Na$\rm^{+}$ $>$ K$\rm^{+}$\ 
$>$ Cs$\rm^{+}$\ at T=350~K.\ When the\ temperature is\ 
323~K and 350~K,\ the diffusivity of Na$\rm^{+}$\ ion\ 
is less than Li$\rm^{+}$ ion.\ This implies that the\ 
reactivity of sodium\ ion is also\ higher than that of\ 
lithium ion\ whenever there is\ increase in temperature.\ 
Thus,\ our choice\ of interlayer\ cation also affects\ 
the diffusivity of\ water.\  
Our simulated result is\ compared with another\ 
natural smectite clay\ called montmorilonite\ and\ 
natural hectorite\ clay.\ For montmorilonite clays,\ 
using monovalent\ counterions called~(Na$\rm^{+}$,\ 
Cs$\rm^{+}$)\ in bihydrated\ system,\ the self diffusion\ 
coefficient is\ $D_{sim}^{ions}=3.7-10.0~{\times}\
10^{-6}{cm^2}s^{-1}$~\cite{marry2003microscopic,chang1995computer}.\\
In the confining\ environment\ of nanochannels,\ 
the lateral self\ diffusion\ coefficients\ for\ 
motion parallel\ to the clay\ sheets\ are more\ 
relevant and necessary.\ Due to this,\ we have\ 
calculated the\ lateral diffusion coefficient\ 
of both interlayer\ cation and\ water for all\ 
temperature values\ and types of\ Fluorohectorite\ 
clay.\ The result is\ presented in Table~\ref{tab5}.\
According to a literature~\cite{malikova2005diffusion},\ 
the lateral diffusion\ coefficient of water\ in\ 
Na-montmorilonite is $D_{xy}=[7.7~{\pm}~1.0]~{\times}\
10^{-6}{cm^2} s^{-1}$~(NSE).\ 
Generally,\ in all cases,\ the self diffusion\ 
coefficient\ of both cations\ and water depend\ 
on temperature.\ As the\ temperature increases,\ 
the diffusion coefficients\ also increase.\ 
Figure~\ref{fig:7}\ shows\ the of MSD of ions\ 
and water at\ different temperatures.\ 
The main factor\ for the retarding\ diffusion\ 
coefficient of interlayer\ water and cation\ is\ 
the electrostatic interaction\ between water\ and\ 
cation with the clay sheet.\ 
%..............Table 4................... 
\begin{table*}[!htbp]
\addtolength{\tabcolsep}{0.001mm}
\renewcommand{\arraystretch}{1.5}
\centering 
\caption{Region-Specific Diffusivity (${\times}10^{-5} cm^2 s^{-1}$) 
         of interlayer cation and water through our Model Synthetic 
         Clay.\label{tab4}}
\begin{tabular}{|l|l|l|l|l|l|l|l|l|}
\hline
{T(K)} & $D_{Cs}^{CsFht}$&$D_{H_2O}^{CsFht}$ & $D_{K}^{KFht}$& $D_{H_2O}^{KFht}$& $D_{Li}^{LiFht}$& $D_{H_2O}^{LiFht}$  & $D_{Na}^{NaFht}$& $D_{H_2O}^{NaFht}$ \\ \hline
293  & 0.11~$\pm$~0.05&1.28~$\pm$~ 0.01 & 0.19~$\pm$~0.01& 1.55~$\pm$~ 0.08& 0.26~$\pm$~0.01&1.12~$\pm$~ 0.09& 0.34~$\pm$~0.01& 1.06~$\pm$~ 0.02 \\  
300  & 0.13~$\pm$~0.09&1.52~$\pm$~ 0.18& 0.29~$\pm$~0.06 & 1.88~$\pm$~ 0.07& 0.48~$\pm$~0.06 &1.17~$\pm$~ 0.18& 0.49~$\pm$~0.07& 1.20~$\pm$~ 0.06 \\ 
323  & 0.11~$\pm$~0.01&2.42~$\pm$~ 0.09& 0.29~$\pm$~0.15& 2.33~$\pm$~ 0.19& 0.69~$\pm$~0.03&1.82~$\pm$~ 0.04& 0.68~$\pm$~0.18& 1.67~$\pm$~ 0.06\\ 
350  & 0.22~$\pm$~0.01&2.97~$\pm$~ 0.08& 0.43~$\pm$~0.23& 2.87~$\pm$~ 0.17& 0.91~$\pm$~0.07&2.59~$\pm$~ 0.22& 0.68~$\pm$~0.08 & 2.45~$\pm$~ 0.10\\
\hline 
\end{tabular}
\end{table*}
%..............Table 5...................
\begin{table*}[!htbp]
\addtolength{\tabcolsep}{0.001mm}
\renewcommand{\arraystretch}{1.5}
\centering 
\caption{Region-Specific Lateral Diffusivity 
         (${\times}10^{-5} cm^2 s^{-1}$) 
         of interlayer cations and water through our Model 
         Synthetic Clay.\label{tab5}}
\begin{tabular}{|l|l|l|l|l|l|l|l|l|}
\hline
{T(K)} & $D_{||}^{Cs}$  &$D_{||(H_2O)}^{CsFht}$ & $D_{||}^{K}$ & $D_{||(H_2O)}^{KFht}$ & $D_{||}^{Li}$&$D_{||(H_2O)}^{LiFht}$ & $D_{||}^{Na}$& $D_{||(H_2O)}^{NaFht}$ \\ \hline
293  & 0.15~$\pm$~ 0.07&1.92~$\pm$~0.02 & 0.27~$\pm$~0.01& 2.32~$\pm$~0.13 & 0.39~$\pm$~0.02&1.68~$\pm$~0.14& 0.50~$\pm$~ 0.01&1.86~$\pm$~ 0.03\\ 
300  & 0.17~$\pm$~0.12&2.28~$\pm$~0.26 & 0.42~$\pm$~0.10&2.83~$\pm$~0.11 & 0.71~$\pm$~0.08&1.75~$\pm$~0.27 & 0.73~$\pm$~0.12&1.80~$\pm$~ 0.103\\ 
323  & 0.15~$\pm$~0.02&2.28~$\pm$~0.26 & 0.42~$\pm$~0.18&3.49~$\pm$~0.29& 1.03~$\pm$~0.05&2.72~$\pm$~0.06 & 1.01~$\pm$~0.28&2.49~$\pm$~ 0.09\\ 
350  & 0.31~$\pm$~0.03 &4.46~$\pm$~0.1 & 0.64~$\pm$~0.33&4.30~$\pm$~0.26 & 01.34~$\pm$~0.10&3.88~$\pm$~0.33& 1.01~$\pm$~ 0.12&3.68~$\pm$~ 0.16\\ 
\hline
\end{tabular}
\end{table*}  
%.............Figure 8.................
\begin{figure*}[htp!]
{\includegraphics[scale=0.5]{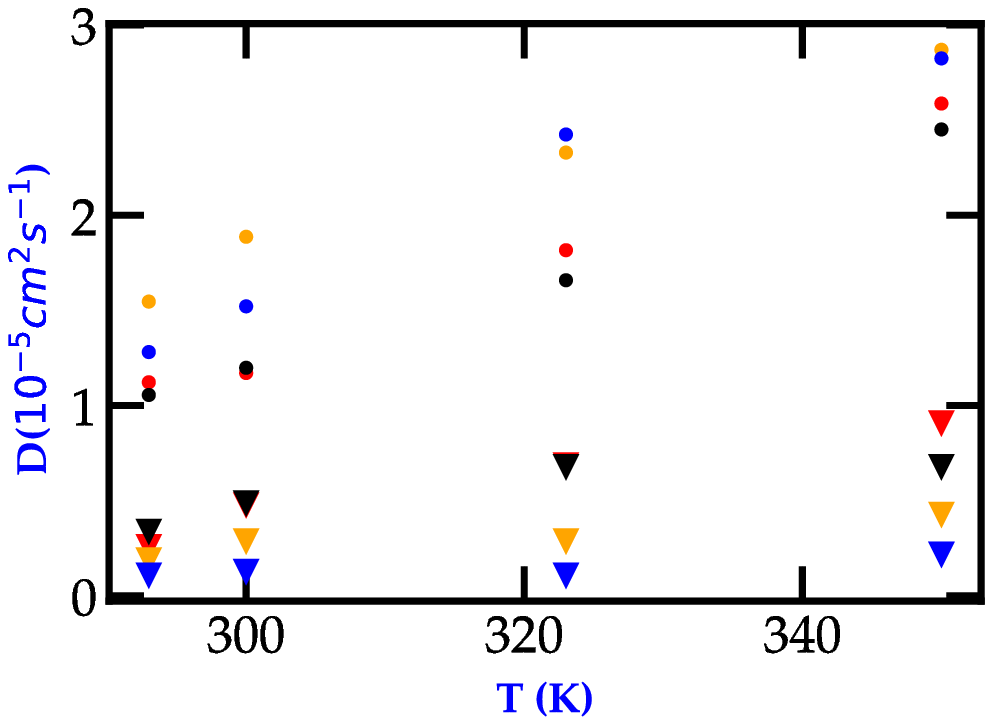}}
{\includegraphics[scale=0.5]{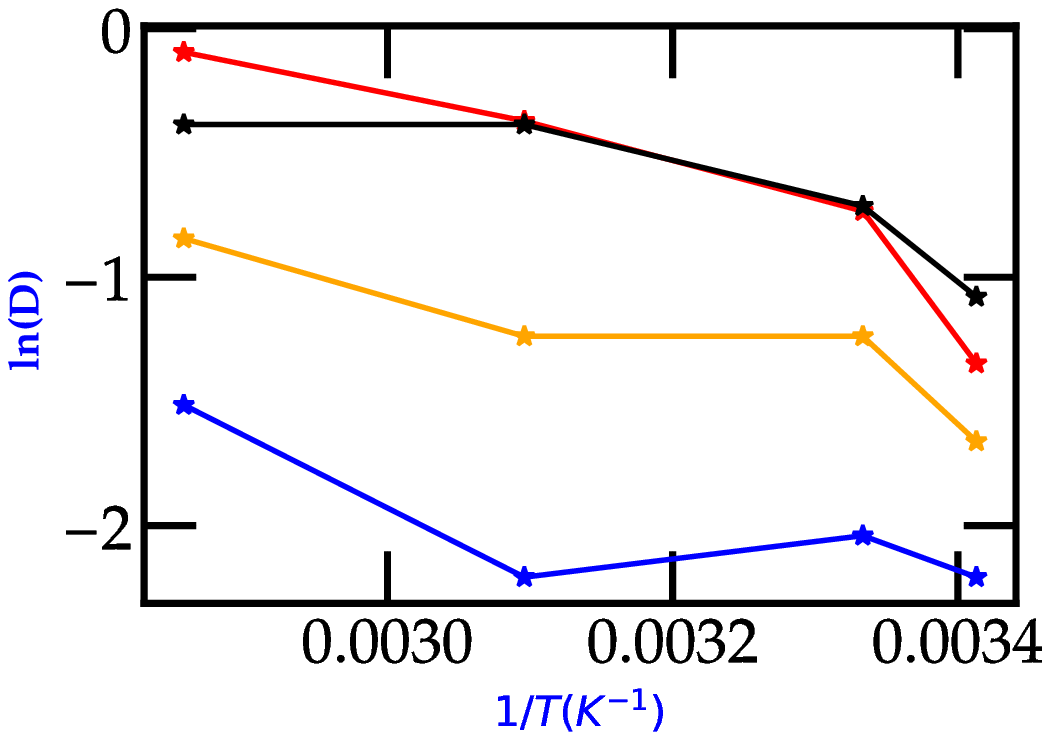}}
{\includegraphics[scale=0.5]{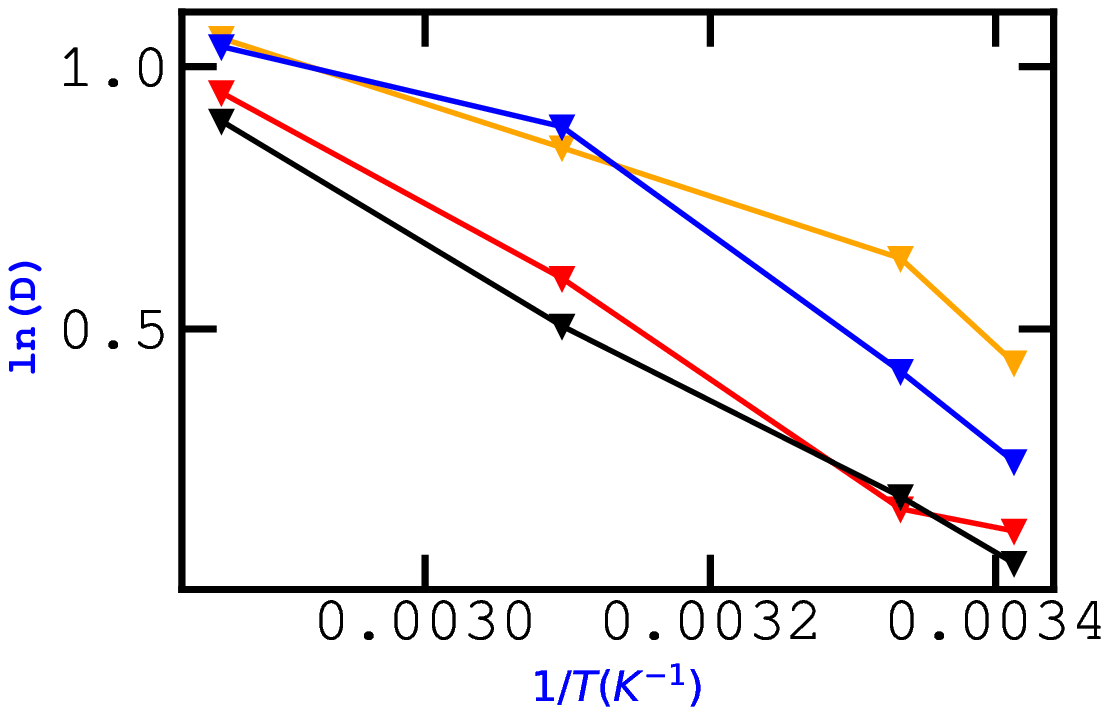}}
\caption{The (a)~Diffusion coefficient vs temperature graph\ 
          and (b) The natural logarithm of 3D diffusion coefficient 
          vs inverse\ temperature graph of\ interlayer molecules of\ 
          Li,\ Na,\ K and Cs-Fht clays.
          (c) The natural logarithm of 3D diffusion coefficient 
          vs inverse\ temperature graph of\ water molecule.
          Colors:~red-LiFht,\ 
          black-NaFht,\ orange-KFht~$\&$~black-CsFht.\label{fig:8}}
\end{figure*} 
%.............Figure 9.................
\begin{figure*}[htp!]
{\includegraphics[scale=0.5]{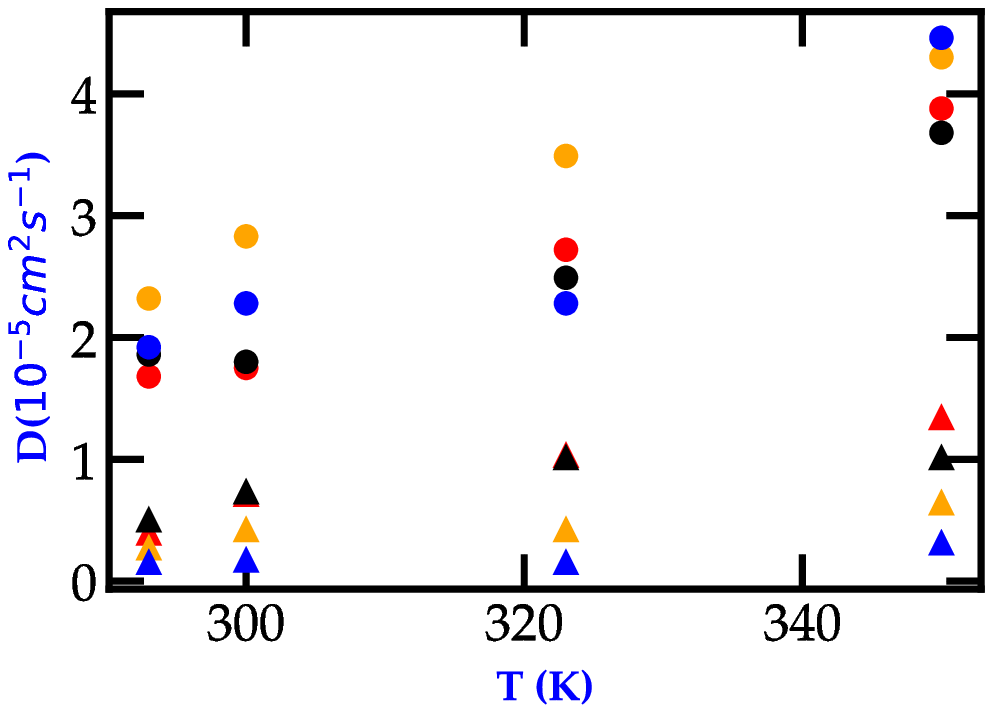}}
{\includegraphics[scale=0.5]{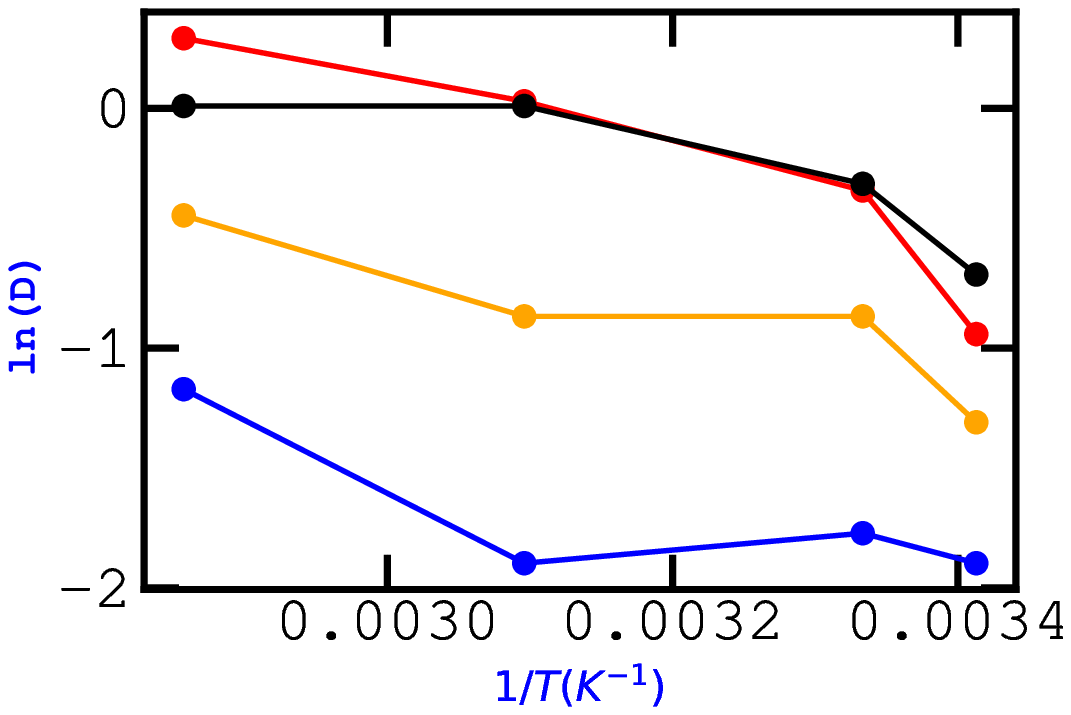}}
{\includegraphics[scale=0.5]{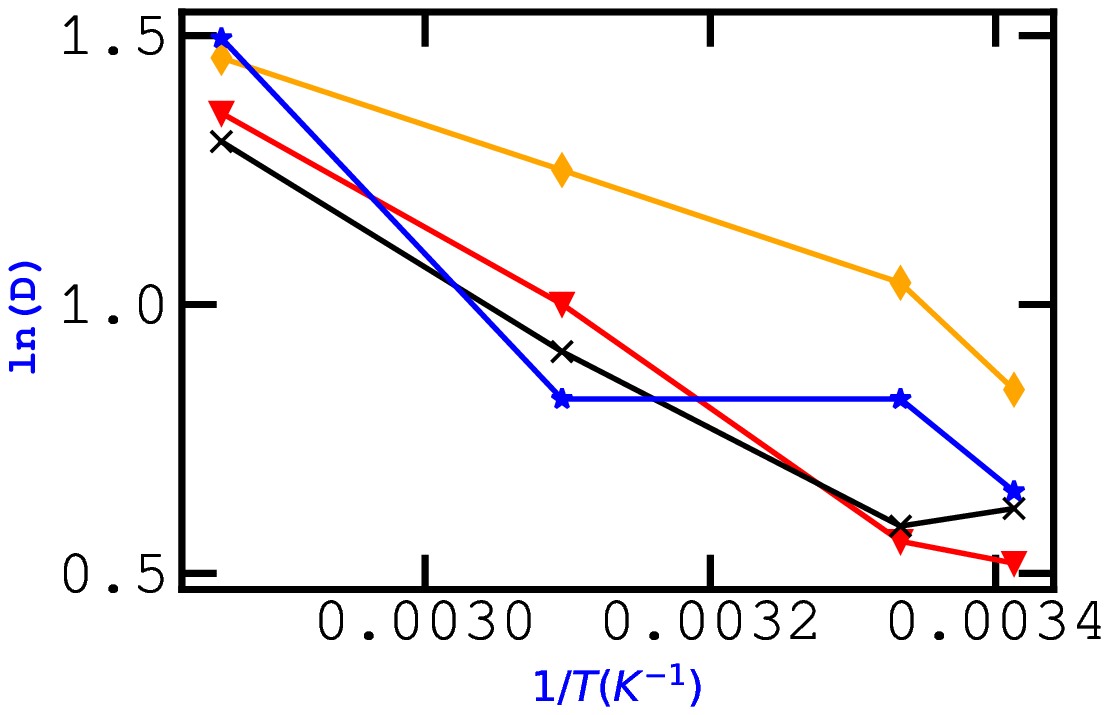}}
\caption{(a)~Lateral diffusion coefficient vs temperature graph\ 
          and (b) The natural logarithm of 2D diffusion coefficient 
          vs inverse\ temperature graph of\ interlayer molecules of\ 
          Li,\ Na,\ K and Cs-Fht clays.
          (c) The natural logarithm of 2D diffusion coefficient 
          vs inverse\ temperature graph of\ water molecule.
          Colors:~red-LiFht,\ 
          black-NaFht,\ orange-KFht~$\&$~black-CsFht.\label{fig:9}} 
\end{figure*} 
Fig.~\ref{fig:8}(b)~$\&$~(c) $\&$ Fig.~\ref{fig:9}(b)~$\&$~(c)\ 
shows the Arrhenius plot of the diffusion\ coefficient\ 
plot of interlayer molecules.\ We may determine\ 
the pre-exponential factor\ by extrapolating the\ 
graph to zero.\ The activation\ energy of water\ 
calculated from\ the simulation was\ 13.05 $\frac{kJ}{mol}$,\ 
12.46~$\frac{kJ}{mol}$,\ 8.68~$\frac{kJ}{mol}$,\ 
$\&$\ 12.07~$\frac{kJ}{mol}$\ in Li,\ Na,\ K,\ and\ 
Cs-Fht,\ respectively,\ and the activation energy\ 
of Li,\ Na,\ K,\ and Cs\ was 16.62~$\frac{kJ}{mol}$,\ 
9.55~$\frac{kJ}{mol}$,\ 9.94~$\frac{kJ}{mol}$,\ 
$\&$~8.52~$\frac{kJ}{mol}$,\ respectively,\ in 3D\ 
motion of molecules.\
\section{Conclusion\label{sec:conc}}
The transport characteristics\ of cations\ 
and water molecules\ in bihydrated Li, Na,\ 
K,\ and Cs-Fht\ clays were studied\ 
using MD simulations. Na$\rm^{+}$\ 
and\ Li$\rm^{+}$ ions exhibit\ a qualitative\ 
difference\ from K$\rm^{+}$ and Cs$\rm^{+}$\ 
ions,\ according to\ trajectory maps\ of the\ 
cations \ on the\ simulation\ time scale 1000 $ps$.\ 
The former\ exhibits significant\ diffusion\ 
motion,\ including hopping events,\ whereas,\ 
the later\ exhibits more\ constrained motion,\ 
probably\ due to its stronger\ interaction\ 
with\ the stiff clay\ surface than with the\ 
mobile\ water molecules.\\
The reason looks to be due to\ more K$\rm^{+}$~$\&$~Cs$\rm^{+}$\ 
ions than Na$\rm^{+}$\ $\&$\ Li$\rm^{+}$ ions adsorbing\ 
on\ the clay surface,\ according\ to density\ 
profiles.\ Consequently,\ it can\ be inferred\ 
that\ K$\rm^{+}$ $\&$ Cs$\rm^{+}$\ ions screen\ 
the\ negatively charged\ surface more\ efficiently\ 
than Na$\rm^{+}$ $\&$\ Li$\rm^{+}$ ions.\
The self-diffusion\ coefficient's\ measurement\ 
reveal that the\ values rise as\ the temperature\ 
rises.\ 
Diffusion\ coefficients\ for bilayer states are\ 
typically in\ the order of $10^{-5} cm^{2} s^{-1}$\ 
in most neutron\ scattering\ experiments\ (mostly TOF)\ 
on natural clays,\ like montmorillonite, hectorite,\ 
or vermiculite.\ In the synthetic hectorite clay,\ 
it is reasonable\ to expect that\ the liquid phase\ 
is evenly distributed\ throughout various\ interlayers\ 
and may even\ be better ordered.\ As a result,\ 
water would move\ more slowly in\ synthetic hectorite\ 
than in\ other systems. But,\ the outcomes\ for Na,\ 
Li, K,\ and\ Cs-Fht are satisfactorily\ consistent\ 
with\ other simulation\ and experiment\ literatures.\ 

This shows that\ the clay interlayer\ 
spacing is\ a crucial path-way for\ the\ 
transportation\ of ions and water,\ and it\ 
supports\ the accuracy of\ the diffusion\ 
coefficients\ investigated by this work.

The present study\ and its future extensions,\ 
for example to\ the case of the\ different water\ 
model in the clay,\ opens the possibility\ to use\
simulations to\ study detailed processes\ occurring\ 
in clays which\ can not be seen experimentally,\ 
to analyse data\ mixing by several\ dynamics\
such as rotation\ and vibrations,\ and to model\ 
accurately\ more and more\ complex systems\ 
approaching\ the real\ natural clays.\
\section*{Disclosure\ statement}
The authors declare that there is no conflict of interest.

\section*{Acknowledgments}
The authors acknowledge\ the help\ of the\ 
computer lab\ at the Physics\ Department of\ 
Addis Ababa University,\ which is supported\ 
by Uppsala University's\ International\ Science\ 
Program (ISP).\ The\ office\ of\ VPRTT\ of Addis\ 
Ababa\ University\ is also warmly\ appreciated\ 
for\ supporting\ this\ research under a\ grant\ 
number\ TR/035/2021.\ 

ORCID~iDs.~K.N.\ Nigussa.\\
\url{https://orcid.org/0000-0002-0065-4325}.\\
\section*{References}   
\bibliographystyle{elsarticle-num}
\bibliography{refs.bib}
\end{document}